\documentclass[4apaper]{aipproc}
\layoutstyle{8x11single}

\usepackage{amssymb}
\usepackage{amsfonts}
\usepackage{latexsym}
\usepackage{epsfig}
\usepackage{verbatim} 
\usepackage{color}  

\def\bR{\begin{color}{red}}
\def\bB{\begin{color}{blue}}
\def\bM{\begin{color}{magenta}}
\def\bC{\begin{color}{cyan}}
\def\bW{\begin{color}{white}}
\def\bBl{\begin{color}{black}}
\def\bG{\begin{color}{green}}
\def\bY{\begin{color}{yellow}}
\def\e{\end{color}}

\def\PP{{\rm P}}

\input{diagrams.sty}
\newarrow{Toto}<--->
\newarrow{Is}=====
\newarrow{Dotsto}....>
\newarrow{Dots}.....

\newcommand{\bit}{\begin{itemize}}
\newcommand{\eit}{\end{itemize}\par\noindent} 
\newcommand{\ben}{\begin{enumerate}}
\newcommand{\een}{\end{enumerate}\par\noindent}
\newcommand{\beq}{\begin{equation}}
\newcommand{\eeq}{\end{equation}\par\noindent}
\newcommand{\beqa}{\begin{eqnarray*}}
\newcommand{\eeqa}{\end{eqnarray*}\par\noindent}
\newcommand{\beqn}{\begin{eqnarray}}
\newcommand{\eeqn}{\end{eqnarray}\par\noindent}

\begin{document}
\title[Kindergarten Quantum Mechanics]{Kindergarten Quantum Mechanics\\ {\rm--- lecture notes ---}}
\author{Bob Coecke}{address={Oxford University Computing Laboratory,\\ Wolfson Building, Parks rd, OX1 3QD Oxford, UK.\\ {\tt coecke@comlab.ox.ac.uk}}}
\date{}
\begin{abstract}
These lecture notes survey some joint work with Samson Abramsky as it was presented by me at several conferences in the summer of 2005. It concerns `doing quantum mechanics using only pictures of lines, squares, triangles and diamonds'.  This picture calculus can be seen as a very substantial extension of Dirac's notation, and has a purely algebraic counterpart in terms of so-called Strongly Compact Closed Categories (introduced by Abramsky and I in \cite{AC1,AC1.5}) which subsumes my Logic of Entanglement \cite{Coe}. For a survey on the `what', the `why' and the `hows' I refer to a previous set of lecture notes \cite{LN1,LN2}.  In a last section we provide some pointers to the body of technical literature on the subject.
\end{abstract}
\maketitle
\parskip .15cm

\section{1. The Challenge}

Why did discovering quantum teleportation take 60 year?
We claim that this is due to a  `bad quantum formalism' (bad $\not=$ wrong) and this badness is in particular due to the fact that the formalism is `too low level' cf. 
\[ 
{\mbox{``GOOD QM''}\over{\tt von\ Neumann\ QM}}
\ \simeq\  
{\rm HIGH\mbox{\rm-}LEVEL\
language\over{\tt low\mbox{\rm-}level\
language}}\,.
\]
%This comparison can indeed be taken quite literally. At the lowest level the state of a computer is a string of zeros and ones, and a program is something that transforms such a state into another one.  If we take such a program to be a relation which relates each input bit to none i.e.~a $\{0,1\}$-valued matrix, we obtain matrix calculus as our programming language theory.  And the Hilbert space formalism is merely a variant on this where we replace the booleans $\mathbb{B}$ by the complex numbers $\mathbb{C}$.  ... we have no itteration, conditionals and all that here, and we have a semiring calculus (boolean logic), not the field calculus, ...so it's quite a limited programing language 
Interestingly enough during one of my talks Gilles Brassard (one of the fathers of teleportation)  disputed my claim on why teleportation was only discovered in the 1990's. He argued that the reason `they' only came up with teleportation when they did was due to the fact that the question had never been asked before \cite{talk} --- and he added that once the question was asked the answer came quite easily (in a couple of hours).  But that exactly confirms my claim: the badness of the quantum formalism causes the question not to be asked!
Moreover, what is a more compelling argument for the badness of a formalism than having its creator on your side?  While von Neumann designed Hilbert space quantum mechanics in 1932 \cite{vN} he renounced it 3 years later \cite{Birk,Redei}:  ``I would like to make a confession which may seem immoral: I do not believe absolutely in Hilbert space no more.'' (sic.)
\par\vspace{-1mm}\noindent
\bit
\item
So, wouldn't it be nice to have a `good' formalism, in which discovering teleportation would be trivial? 
\item I claim that such a formalism already exist! That's what these notes are all about!
\item So you think it must be absurdly abstract coming from guys like us? 
\item Not at all! In fact, it could be taught in kindergarten!
\eit

\section{2. Category Theory}

Of course we do not expect you to know about category theory, nor do we want to encourage you here to do so.  The whole point about these notes is the graphical calculus which could be taught in kindergarten.  But we just want to mention that what is really going on `behind the scene' is category theory, although you won't notice it.  Below we give a simple but extremely compelling argument why category theory does arise very naturally in physics. 

Why would a physicist care about category theory, why would he want to know about it, why would he want to show off with it?  There could be many reasons.  For example, you might find John Baez's webside one of the coolest in the world.  Or you might be fascinated by Chris Isham's and Lee Smolin's ideas on the use of topos theory in Quantum Gravity.  Also the connections between knot theory, braided categories, and sophisticated mathematical physics might lure you, or you might be into topological quantum field theory. Or, if you are also into pure mathematics,  you might just appreciate category theory due to its \em unifying power of mathematical structures and constructions\em.  But there is a far more obvious reason which is never mentioned.  Namely, \em a category is the exact mathematical structure of practicing physics\em!  
What do I mean here by practicing physics?  You take a physical system of type $A$ (e.g.~a qubit, or two qubits, or an electron) and perform an operation $f$ on it (e.g.~perform a measurement on it) which results in a system possibly of a different type $B$ (e.g.~the system together with classical data which encodes the measurement outcome, or, just classical data in the case that the measurement destroyed the system). So typically we have 
\begin{diagram}
A&\rTo^{f}&B
\end{diagram}
where $A$ is the initial type of the system, $B$ is the resulting type and $f$ is the operation. One can perform an operation 
\begin{diagram}
B&\rTo^{g}&C
\end{diagram}
after $f$
since the resulting type $B$ of $f$ is also the initial type of $g$, and we write $g\circ f$ for the consecutive application of these two operations. Clearly we have $(h\circ g)\circ f=h\circ(g\circ f)$ since putting the brackets merely adds the superficial data on conceiving two operations as one.  If we further set 
\begin{diagram}
A&\rTo^{1_A}&A
\end{diagram}
for the operation `doing nothing on a system of type $A$' we have $1_B\circ f=f\circ 1_A=f$. Hence we have a category!  Indeed, the (almost) precise definition of a category is the following:
\par\medskip\noindent{\bf Definition.}
A {\em category} ${\bf C}$ consists of:
\bit 
\item \em objects \em $A,B,C, \ldots$\,,
\item  \em morphisms  \em $f,g,h,\ldots \in{\bf C}(A,B)$ for each pair $A,B$\,,
\item  associative \em composition \em i.e.
$f\in{\bf C}(A,B)\ ,\ g\in{\bf C}(B,C)\ \Rightarrow\ g\circ f\in {\bf C}(A,C)$
with $(h\circ g)\circ f=h\circ (g\circ f)$\,,
\item an \em identity morphism \em  $1_A\in{\bf C}(A,A)$ for each $A$ i.e.
$f\circ 1_A=1_B\circ f=f$.
\eit
\par\medskip\noindent
When in addition we want to be able to conceive two systems $A$ and $B$ as one whole $A\otimes B$, and also to consider compound operations $f\otimes g:A\otimes B\to C\otimes D$, then we pass from ordinary categories to a (2-dimensional) variant called monoidal categories (due to Jean Benabou \cite{Benabou}).  For the same \em operational \em reasons as discussed above
category theory could be expected to play an important role in other fields where operations/processes play a central role e.g.~Programing (programs as processes) and Logic \& Proof Theory (proofs as processes), and indeed, in these fields category theory has become quite common practice --- cf.~many available textbooks.
\bR\begin{center}
\begin{tabular}{|c|c|c|} 
\hline
{\rm LOGIC\ \&\ PROOF\ THEORY} & 
{\rm PROGRAMMING} & 
{\rm(OPERATIONAL)\ PHYSICS}\\
\hline
\bBl Propositions\e & \bBl Data Types\e & \bBl Physical System\e\\
\hline
\bBl Proofs\e & \bBl Programs\e & \bBl Physical Operation\e\\
\hline 
\end{tabular}
\end{center}\e
But the amazing thing of the kind of category theory we need here is that it \em formally justifies its own formal absence\em, in the sense that at an highly abstract level you can prove that the abstract algebra is equivalent to merely drawing some pictures (see the last section of these notes for some references on this).  Unfortunately, the standard existing literature on category theory (e.g.~\cite{MacLane}) might not be that suitable for the audience we want to address in this draft.\footnote{Some lecture notes intended for researchers in Foundations of Physics and Quantum Informatics are imminent \cite{CatLN}.}  Category theory literature typically addresses the (broadminded \& modern) pure mathematician and as a consequence the presentations are tailored towards them. The typical examples are various categories of mathematical structures and the main focus is on their similarities in terms of mathematical practice.   The `official' birth of category theory is indeed associated with the Eilenberg-MacLane paper `General Theory of Natural Equivalences' (1945) \cite{EilenbergMacLane} in which the authors observe that the collection of mathematical objects of some given kind/type, when equipped with the maps between them, deserves to be studied in its own right as a mathematical structure since this study entails unification of constructions arising from different mathematical fields such as geometry, algebra, topology, algebraic topology etc.  
 
\section{3. the language of pictures}

In this section we proceed as follows. First we define a language which purely consists of pictures, which has some primitive data (cf.~lines, boxes, triangles and diamonds), in which we have two kinds of composition, namely parallel (conceiving two systems as a compound single one)  and sequential (concatenation in time), and which will obey a certain axiom.  Then we derive some `recognizable' results using this calculus --- e.g.~the picture analogue to teleportation, logic-gate teleportation and entanglement swapping. Finally we show that Hilbert space quantum mechanics is an incarnation of such a picture calculus, so every result we derived in the picture calculus is a proof of some statement about actual quantum mechanics. 

\subsection{3.a. Defining the Calculus}

The primitive data of our formalism consists of  (i) boxes with an input and an output which we call `operation' or `channel', (ii) triangles with only an output which we call `state' or `preparation procedure' or `ket', (iii) triangles with only an input which we call `co-state' or `measurement branch' or `bra', (iv) diamonds without inputs or outputs which we call `values' or `probabilities' or `weights', (v) lines which might carry a symbol to which we refer as the `type' or the `kind of system', and the $A$-labeled line itself will be  conceived as  `doing nothing to a system of type $A$' or the `identity on $A$':
\par\vspace{1mm}\noindent
\begin{minipage}[b]{1\linewidth}
\centering{\epsfig{figure=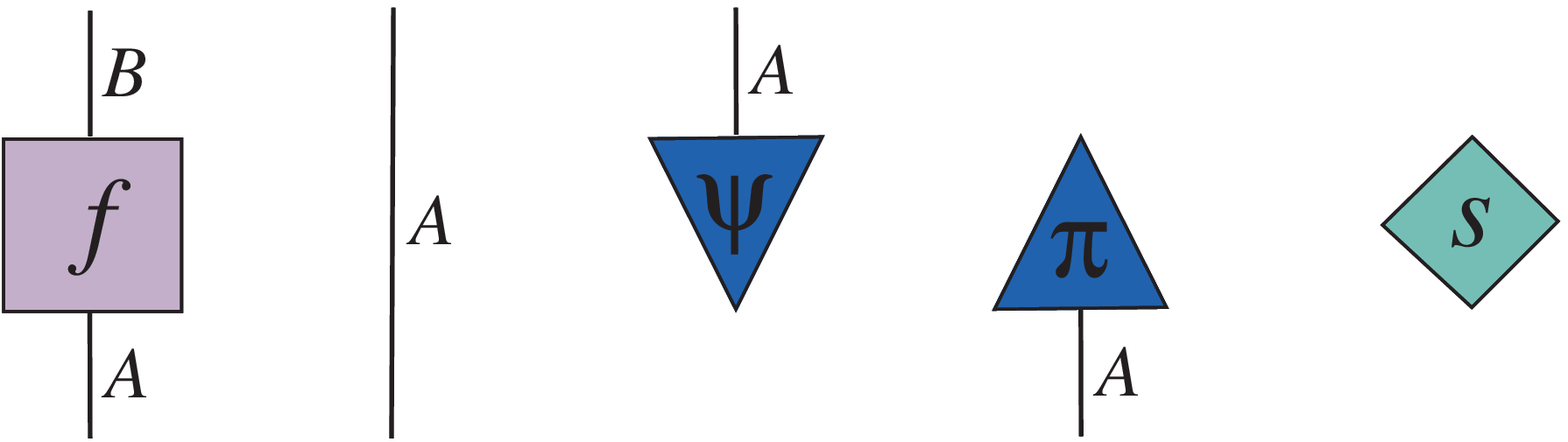,width=192pt}}     
\end{minipage}
\par\vspace{1mm}\noindent
We in particular will think of `no input/output' as a special case of an input/output, so all what applies to square boxes in particular applies to triangles and diamonds.  All what applies to square boxes also applies to the lines.  Parallel and sequential  composition are respectively obtained by (i) placing boxes side by side, and, (ii) connecting up the inputs and outputs of boxes (provided there are any) by lines:
\par\vspace{3mm}\noindent
\begin{minipage}[b]{1\linewidth}
\centering{\epsfig{figure=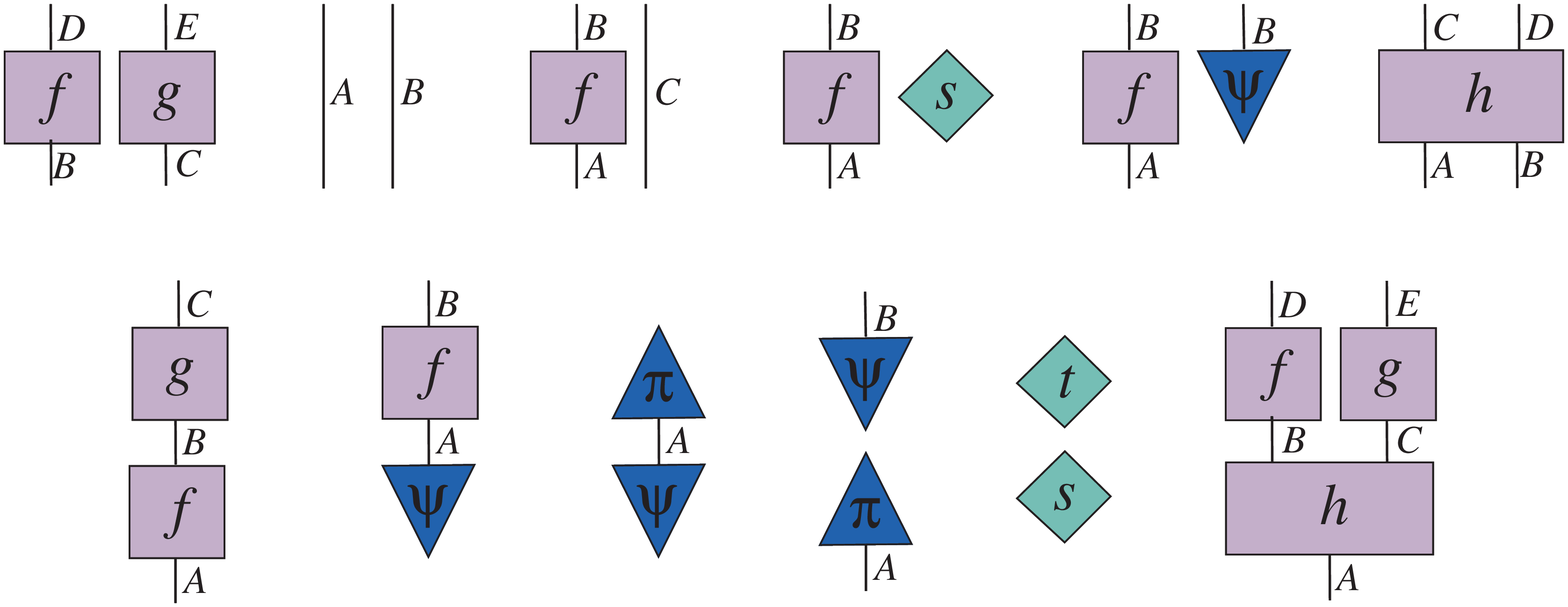,width=337pt}}     
\end{minipage}
\par\vspace{2mm}\noindent
We discuss some interesting special cases.
If we connect up a state and a costate (i.e.~we produce a bra-ket) we obtain a diamond-shape since no inputs nor outputs remain. Thus we obtain what we called a probability.  On the other hand if we connect up a costate and a state (at their input/output-less side i.e. we produce a ket-bra) we obtain a square-shape with a genuine input and a genuine output.
For the input/output-less sides of triangles and diamonds parallel and sequential composition coincide:
\par\vspace{1mm}\noindent
\begin{minipage}[b]{1\linewidth}
\centering{\epsfig{figure=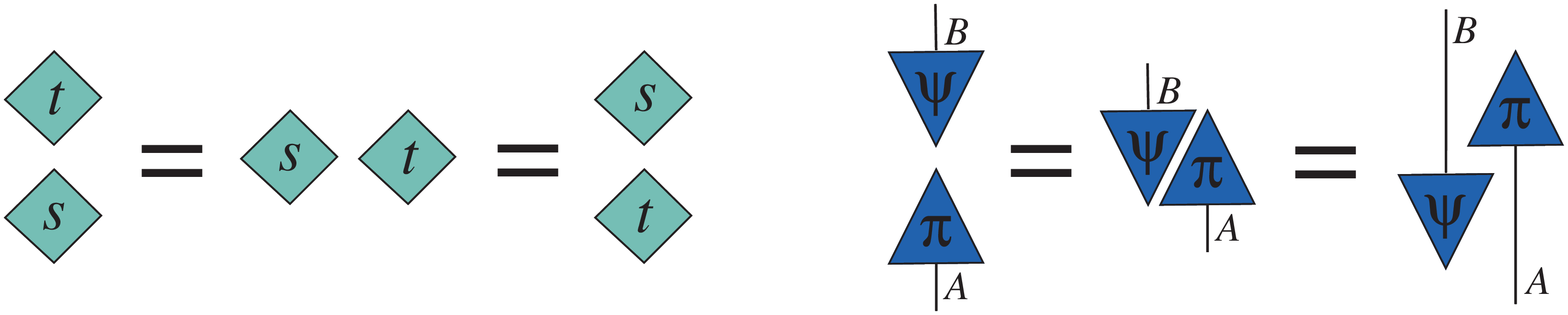,width=337pt}}     
\end{minipage}
\par\vspace{2mm}\noindent
Hence  one could say that input/output-less ends are \em non-local \em in the plane in which the pictures live, and in particular does it follow that sequential composition of diamonds, to which we also will refer as \em multiplication\em, is always \em commutative\em.  For general squares we have the obvious flow-chart-like transformation rules:
\par\vspace{3mm}\noindent
\begin{minipage}[b]{1\linewidth}
\centerline{\epsfig{figure=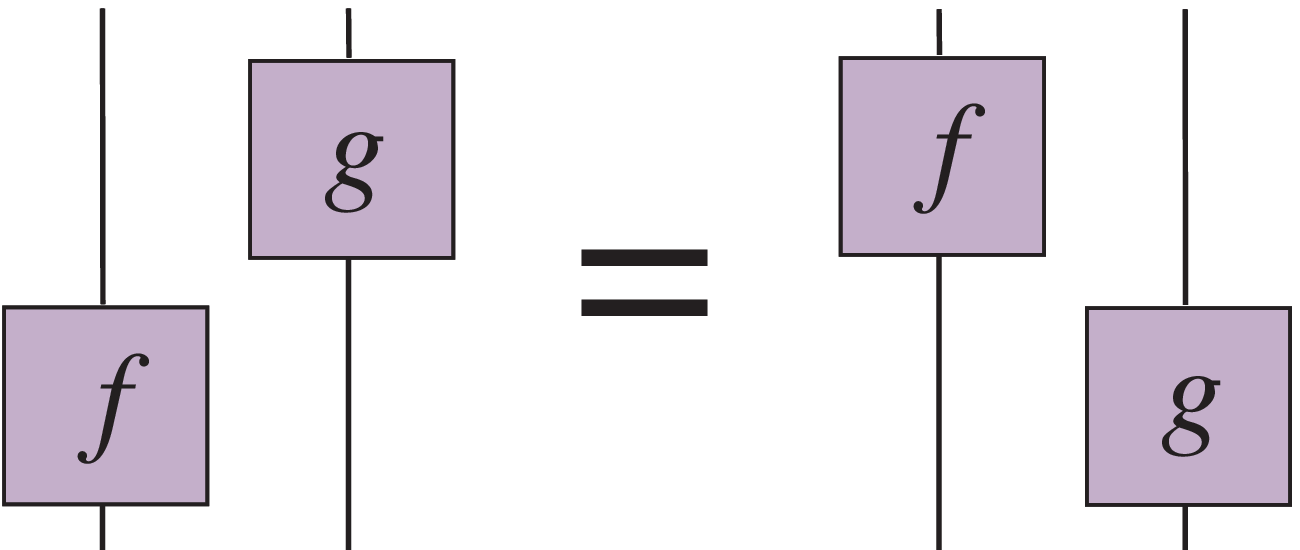,width=144pt}
\qquad\qquad\qquad
\epsfig{figure=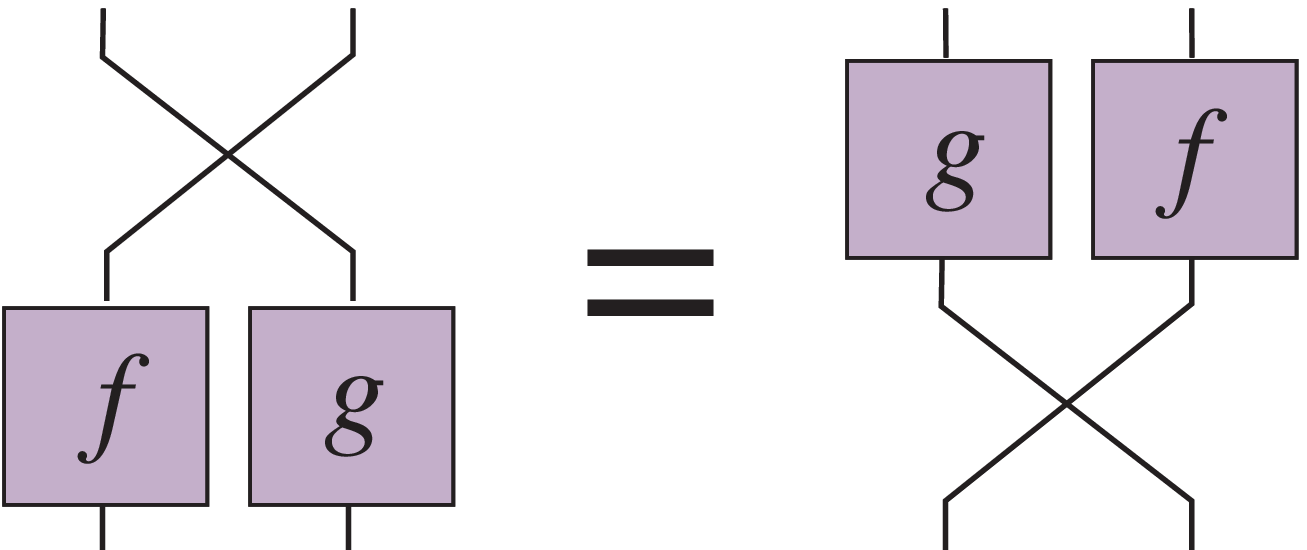,width=144pt}}
\end{minipage}
\par\vspace{3mm}\noindent
Next we are going to add a bit more structure to our pictures.  First we will assume that lines carry an orientation. This means that there exists an operation on types which sends each type $A$ to a type $A^*$ with the opposite orientation. We refer to $A^*$ as  $A$'s \em dual\em, and in particular do we have that $(-)^*$ is an involution i.e.~$(A^*)^*=A$.\footnote{Although related, this should not be thought of as the dual space but as the conjugate space, since otherwise $(-)^*$ would not be an involution.} We also assume that for each box $f:A\to B$ there exists an upside down box $f^\dagger:B\to A$ called $f's$ \em adjoint \em --- note that we do preserve the orientation of the input/output-type.\footnote{This subtle issue where we assume that the adjoint does not alter the orientation of the input/output-types will provide the picture calculus with the full-blown structure of complex conjugation, the notion of unitarity, an inner-product etc. Categorically, this is the feature which `strong' compact closure \cite{AC1,AC1.5} adds as compared to Kelly's ordinary compact closure \cite{Kelly}.} In pictures:
\par\vspace{3mm}\noindent
\begin{minipage}[b]{1\linewidth}
\centering{\epsfig{figure=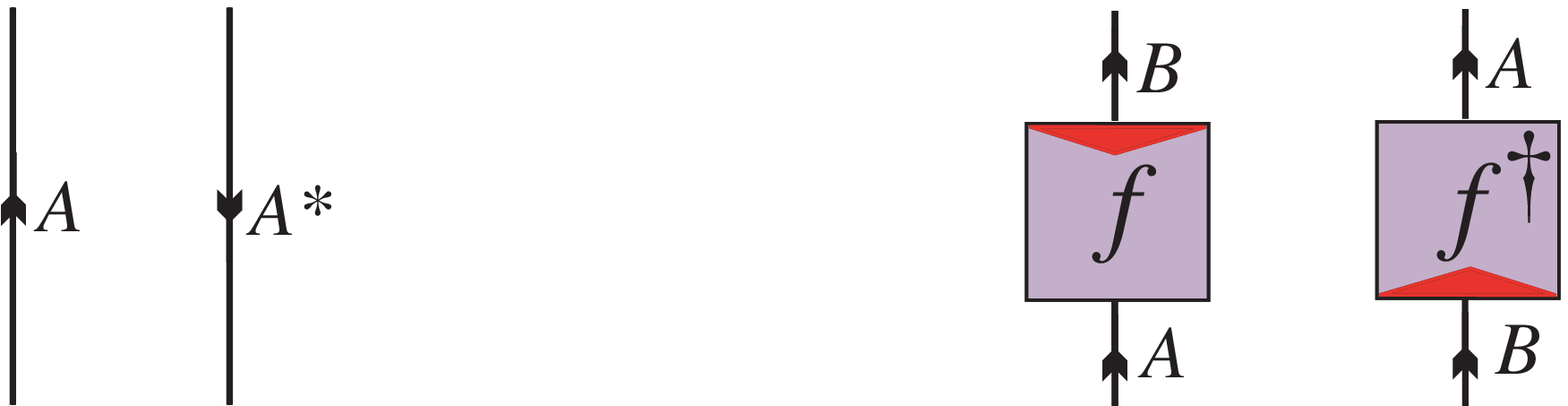,width=200pt}}     
\end{minipage}
\par\vspace{3mm}\noindent
Finally we assume for each type $A$ the existence of a corresponding `Bell-state' or `entanglement-unit':
\par\vspace{3mm}\noindent
\begin{minipage}[b]{1\linewidth}
\centering{\ \epsfig{figure=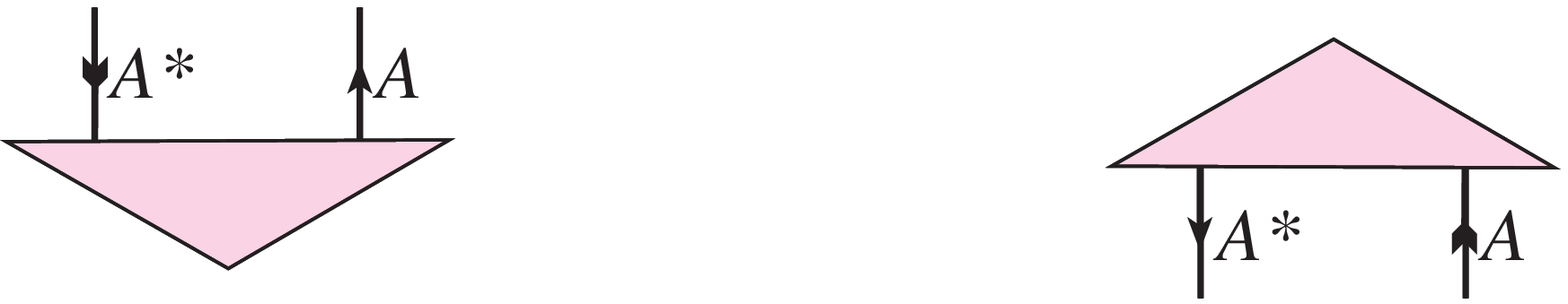,width=205pt}\quad\quad}     
\end{minipage}
\par\vspace{2mm}\noindent
each of which has a `Bell-costate' as its adjoint. The sole axiom we impose is:
\par\vspace{3mm}\noindent
\begin{minipage}[b]{1\linewidth}
\centering{\epsfig{figure=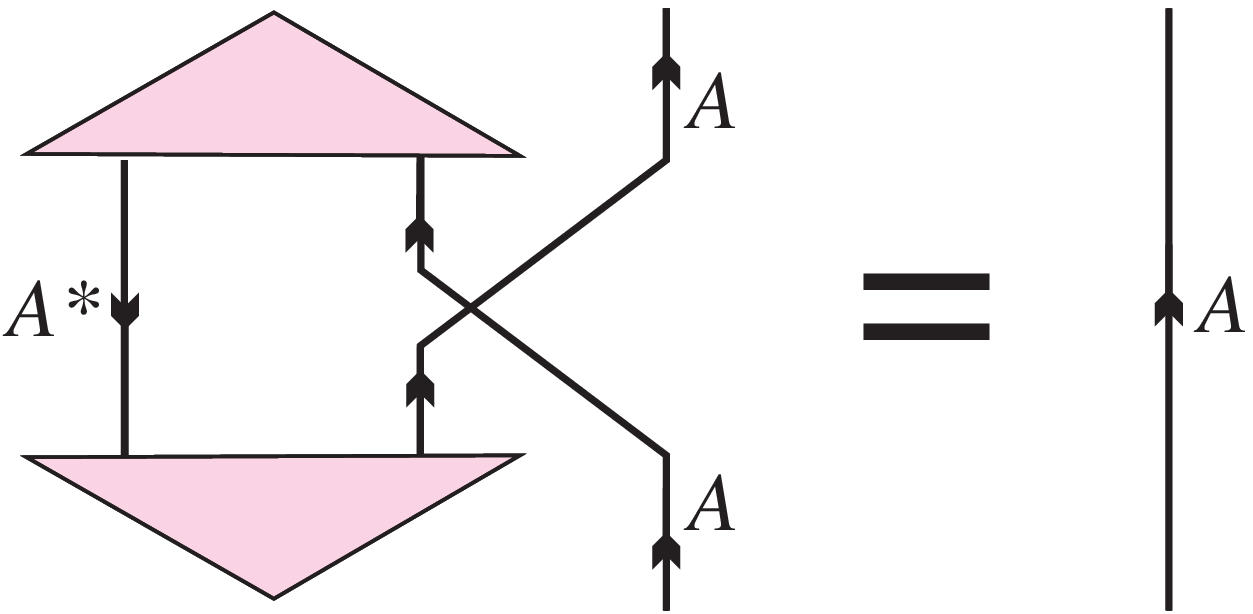,width=144pt}}      
\end{minipage}
\par\vspace{3mm}\noindent
However, if we extend the graphical notation of Bell-states and Bell-costates a bit: 
\par\vspace{3mm}\noindent
\begin{minipage}[b]{1\linewidth}
\centering{\epsfig{figure=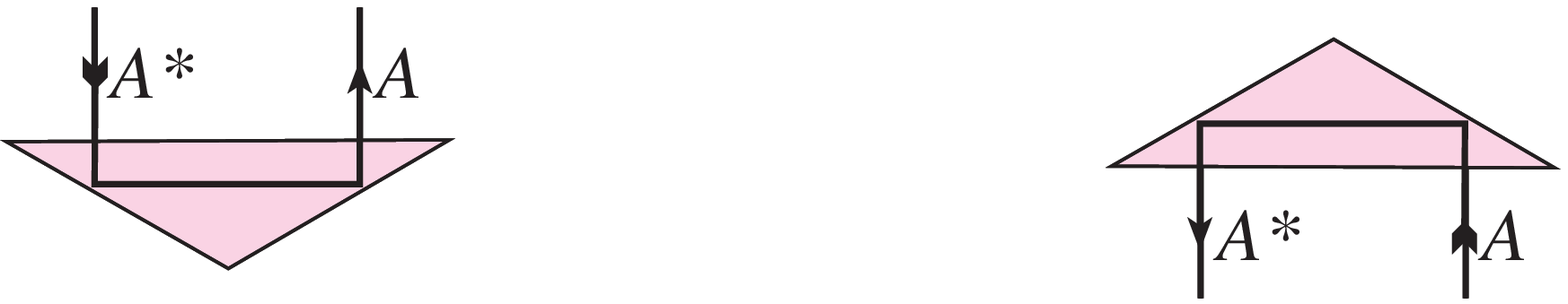,width=205pt}\quad\quad\ }     
\end{minipage}
\par\vspace{1mm}\noindent
we obtain a far more lucid interpretation for the axiom:
\par\vspace{3mm}\noindent
\begin{minipage}[b]{1\linewidth}
\centering{\fbox{\qquad\epsfig{figure=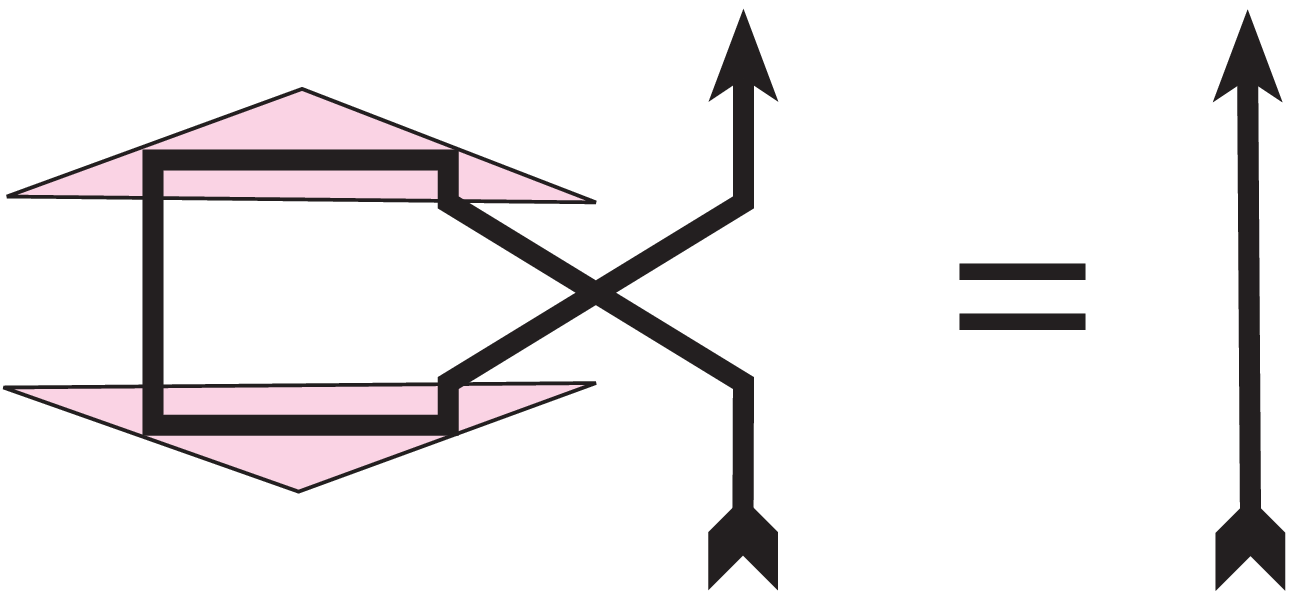,width=200pt}\qquad\hfill{\bf (1)}}}     
\end{minipage}
\par\vspace{3mm}\noindent
The axiom now tells us that we are allowed to \em yank \em the black line:
\par\vspace{1mm}\noindent
\begin{minipage}[b]{1\linewidth}
\centering{\epsfig{figure=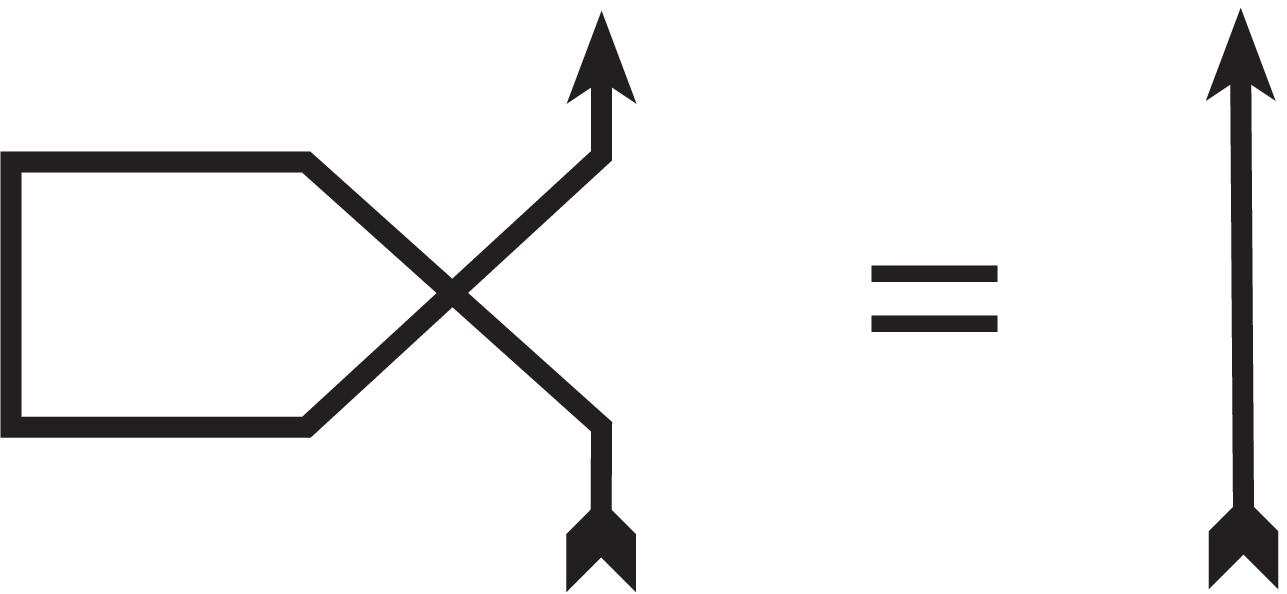,width=160pt}}     
\end{minipage}
\par\vspace{1mm}\noindent
and we call this black line the `quantum information-flow'.  Furthermore,  by simple graph-manipulation we have that:
\par\vspace{1mm}\noindent
\begin{minipage}[b]{1\linewidth}
\centering{\epsfig{figure=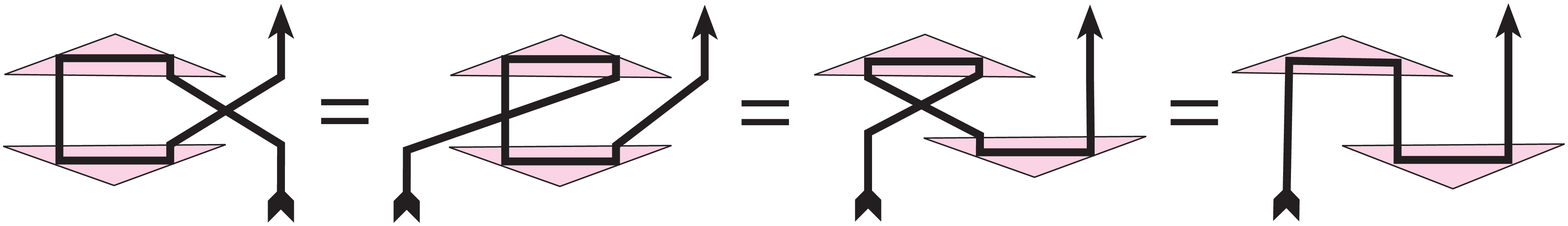,width=460pt}}     
\end{minipage}
\par\vspace{1mm}\noindent
so it follows that the axiom can equivalently be written down as:
\par\vspace{3mm}\noindent
\begin{minipage}[b]{1\linewidth}
\centering{\fbox{\qquad\epsfig{figure=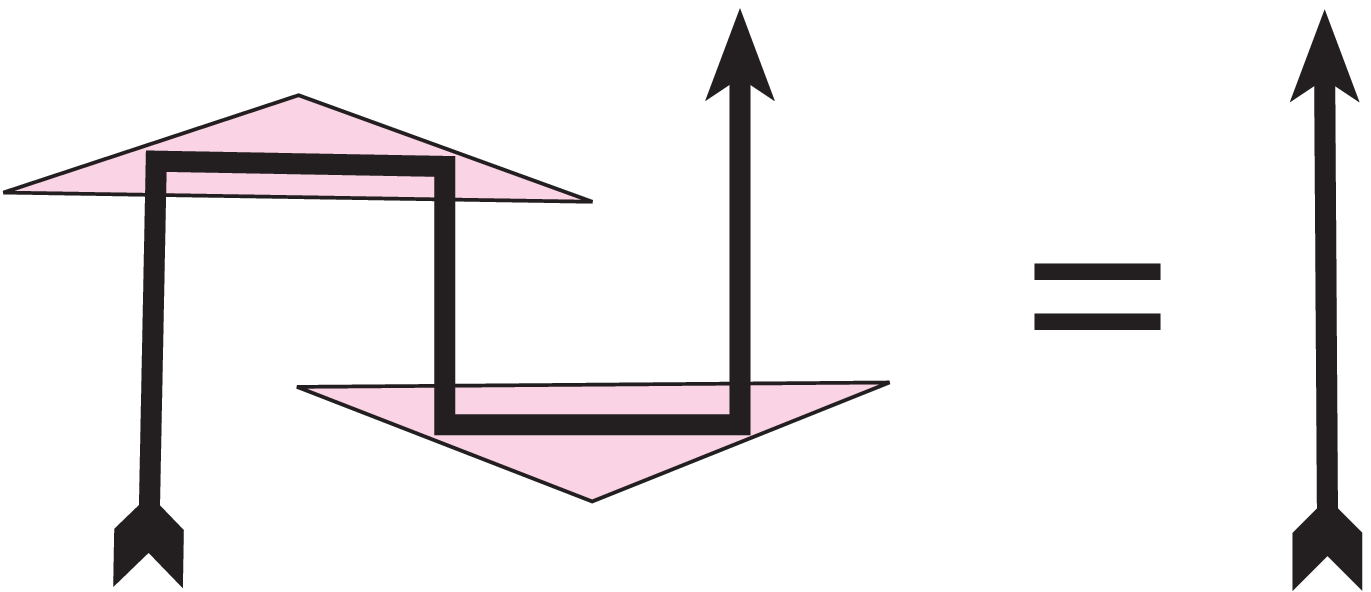,width=200pt}\qquad\hfill{\bf (2)}}}     
\end{minipage}
\par\vspace{3mm}\noindent
which is of course again an instance of yanking:
\par\vspace{1mm}\noindent
\begin{minipage}[b]{1\linewidth}
\centering{\epsfig{figure=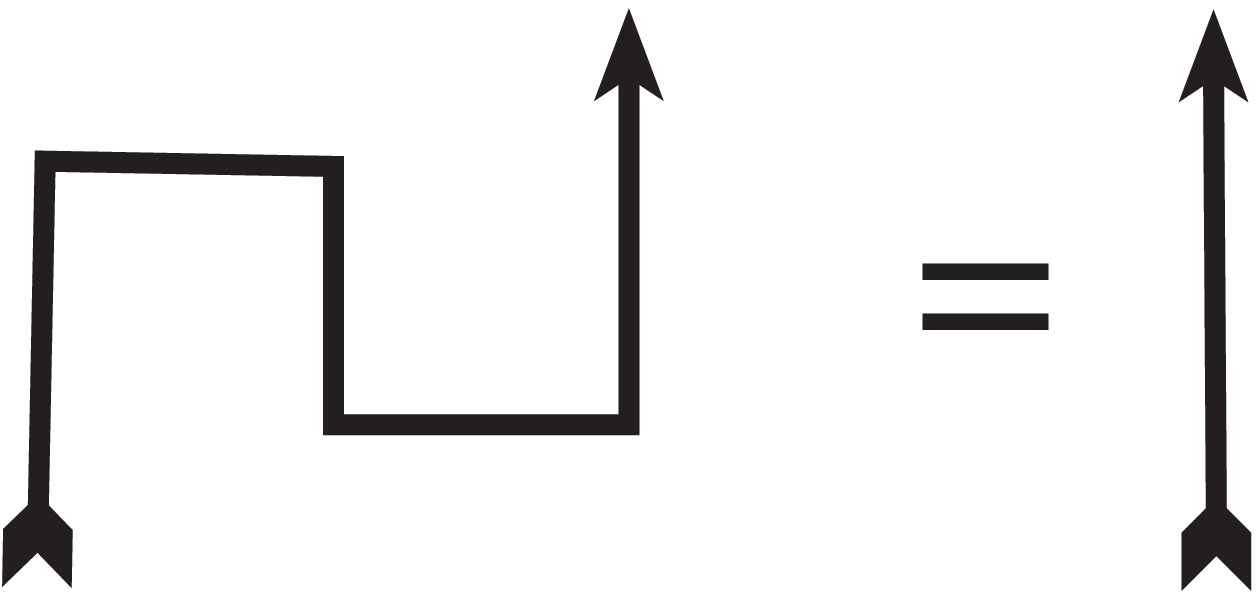,width=160pt}}     
\end{minipage}
\par\vspace{1mm}\noindent
One could even consider dropping the triangles in the notation of Bell-states and Bell-costates but we prefer to keep them since they witness the presence of actual physical entities, and capture the bra/ket-logic in an explicit manner.

\subsection{3.b. Entanglement Compositionality Lemmas} 

We introduce `operation-labeled bipartite (co)states' as:
\par\vspace{3mm}\noindent
\begin{minipage}[b]{1\linewidth}
\centering{\fbox{\quad\epsfig{figure=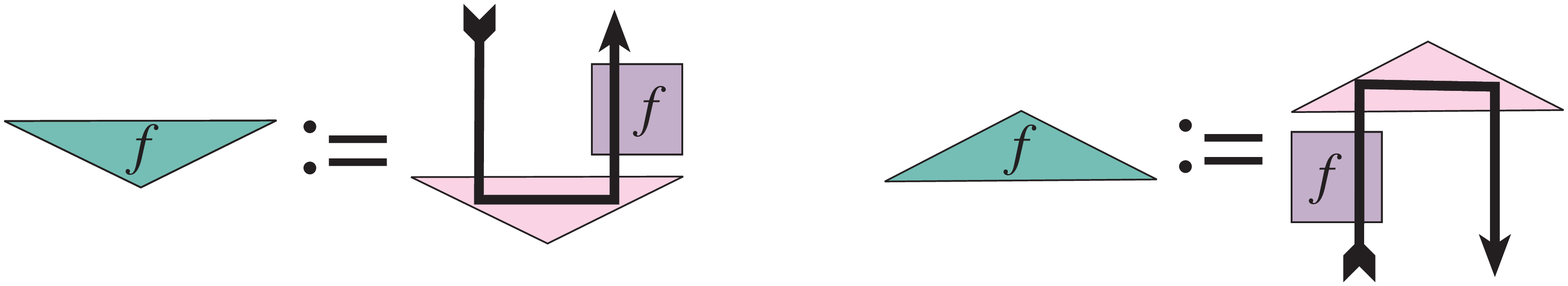,width=420pt}\quad\hfill{\bf (3)}}}     
\end{minipage}
\par\vspace{3mm}\noindent
where the Bell-state for type $A$ is now the special case of a $1_A$-labeled state.
Unfolding definitions we obtain:
\par\vspace{1mm}\noindent
\begin{minipage}[b]{1\linewidth}
\centering{\epsfig{figure=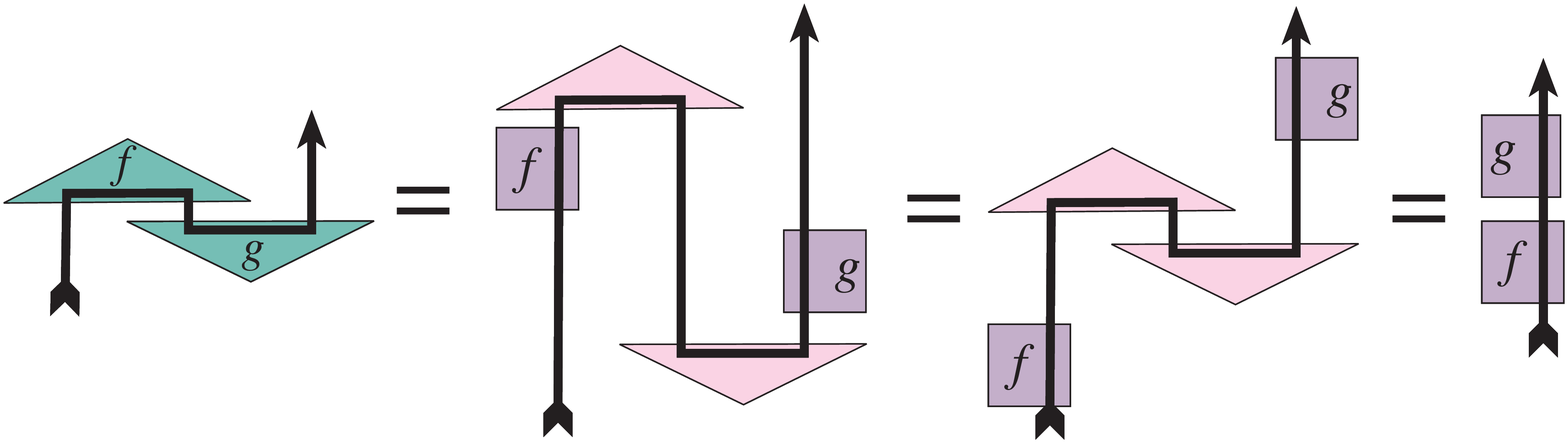,width=400pt}}     
\end{minipage}
\par\vspace{1mm}\noindent
resulting in the following important \em compositionality lemma \em (first proved in \cite{AC1}):
\par\vspace{3mm}\noindent
\begin{minipage}[b]{1\linewidth}
\centering{\fbox{\qquad\epsfig{figure=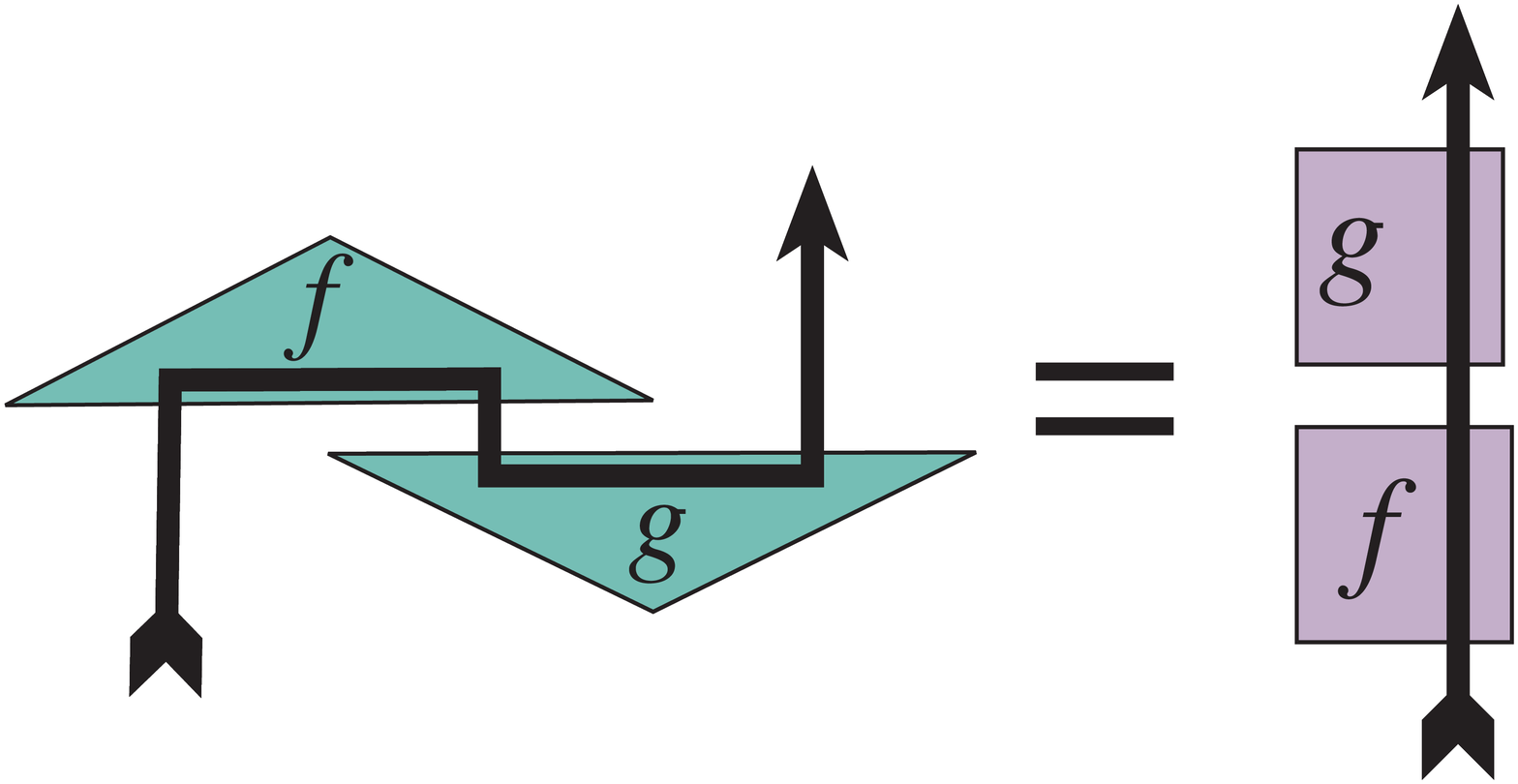,width=200pt}\qquad\hfill{\bf (4)}}}     
\end{minipage}
\par\vspace{3mm}\noindent
and similarly we can derive  (also first proved in \cite{AC1}):
\par\vspace{3mm}\noindent
\begin{minipage}[b]{1\linewidth}
\centering{\fbox{\qquad\epsfig{figure=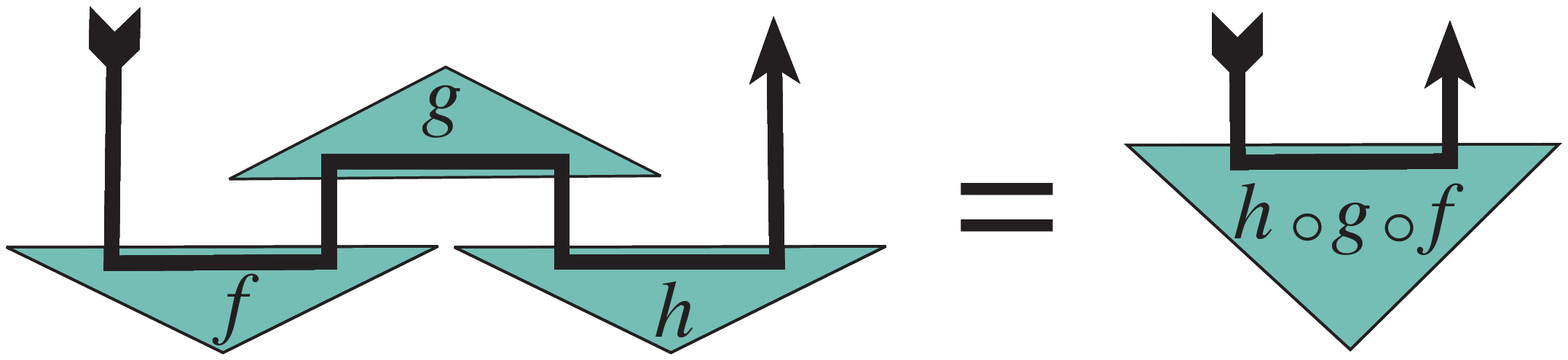,width=260pt}\qquad\hfill{\bf (5)}}}     
\end{minipage}
\par\vspace{3mm}\noindent
Note that we already exposed a quite remarkable feature of the picture calculus.  If we assume `physical time' to flow from bottom to top, then while in picture {\bf (4)} on the left of the equality we \em first \em encounter a $g$-labeled triangle and \em then \em an $f$-labeled triangle, on the right we \em first \em have the $f$-labeled square and only \em then \em the $g$-labeled square. This seems to imply that some weird reversal in the causal order is taking place. A similar phenomenon is exposed in picture {\bf (5)}.
However, a careful logical analysis makes clear that all this can be de-mystified in terms of imposing  (non-commutative) constraints on distributed entities.

\subsection{3.c. Teleportation-like protocols}

The spirit of teleportation is actually already present in the axiom in its form of picture {\bf(2)}.\footnote{Thefact that there was already a presence of teleportation in this axiom was first noted by Abramsky's student Ross Duncan \cite{Duncan}.}  Indeed, we start with a Bell-state of which the first subsystem is in Alice's hands while the second one is in Bob's hands.  Then Alice applies a Bell-costate to the pair consisting of some input state $\psi$ and her part of the Bell-state, resulting in Bob's part of the Bell-state being in state $\psi$:
\par\vspace{3mm}\noindent
\begin{minipage}[b]{1\linewidth}
\centering{\epsfig{figure=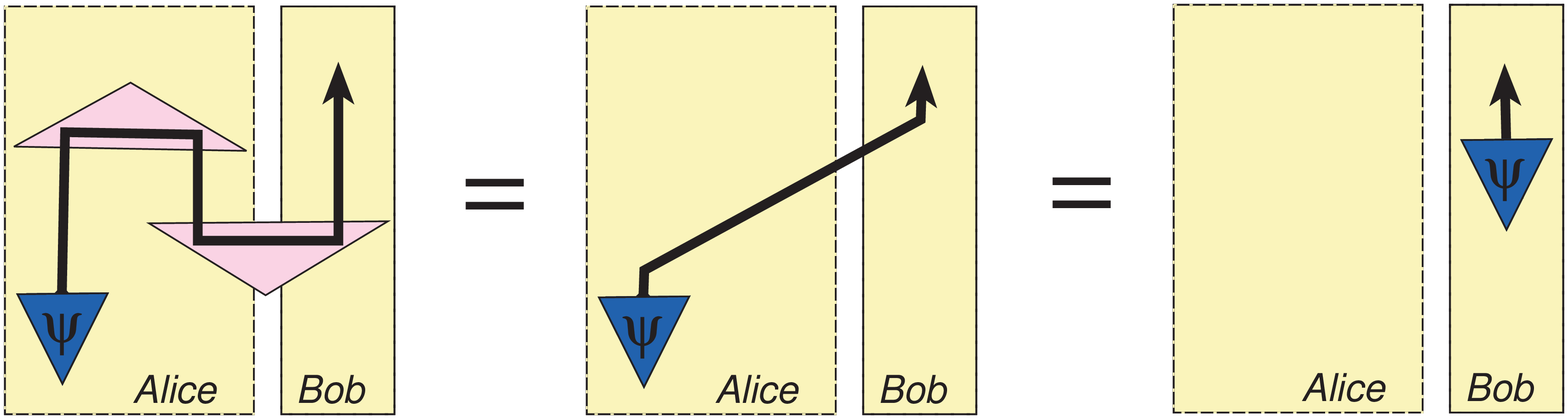,width=400pt}}     
\end{minipage}
\par\vspace{3mm}\noindent
What is still missing here is an interpretation of the costate, and, what will turn out to be related, the fact that we do not seem to need any classical communication.  Within our picture calculus we define a (non-degenarate) \em $f$-labeled bipartite projector \em as the following ket-bra:
\par\vspace{3mm}\noindent
\begin{minipage}[b]{1\linewidth}
\centering{\fbox{\bW{\bf (6)}\e\hfill\qquad\epsfig{figure=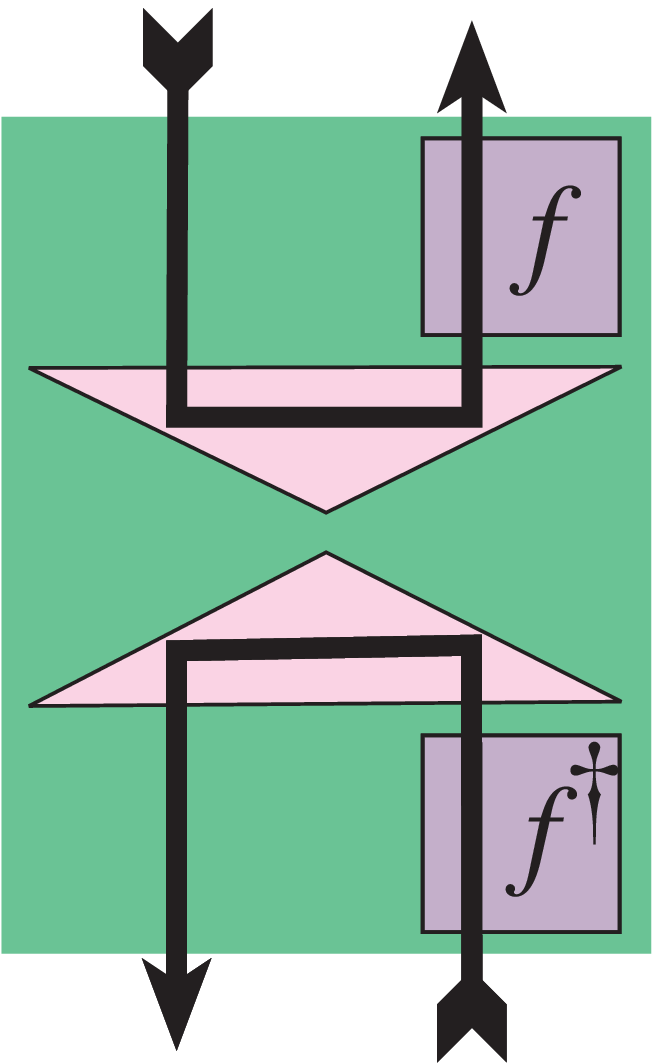,width=80pt}\qquad\hfill{\bf (6)}}}     
\end{minipage}
\par\vspace{3mm}\noindent
of which the  type is ${\rm P}_f:A^*\otimes B\to A^*\otimes B*$ given the  operation $f:A\to B$.  Thus a costate is the  `bottom half' of such a bipartite projector, and we can now think of the teleportation protocol as  involving a Bell-state and an identity-labeled bipartite projector i.e.~a projector which projects on the Bell-state:
\par\vspace{3mm}\noindent
\begin{minipage}[b]{1\linewidth}
\centering{\epsfig{figure=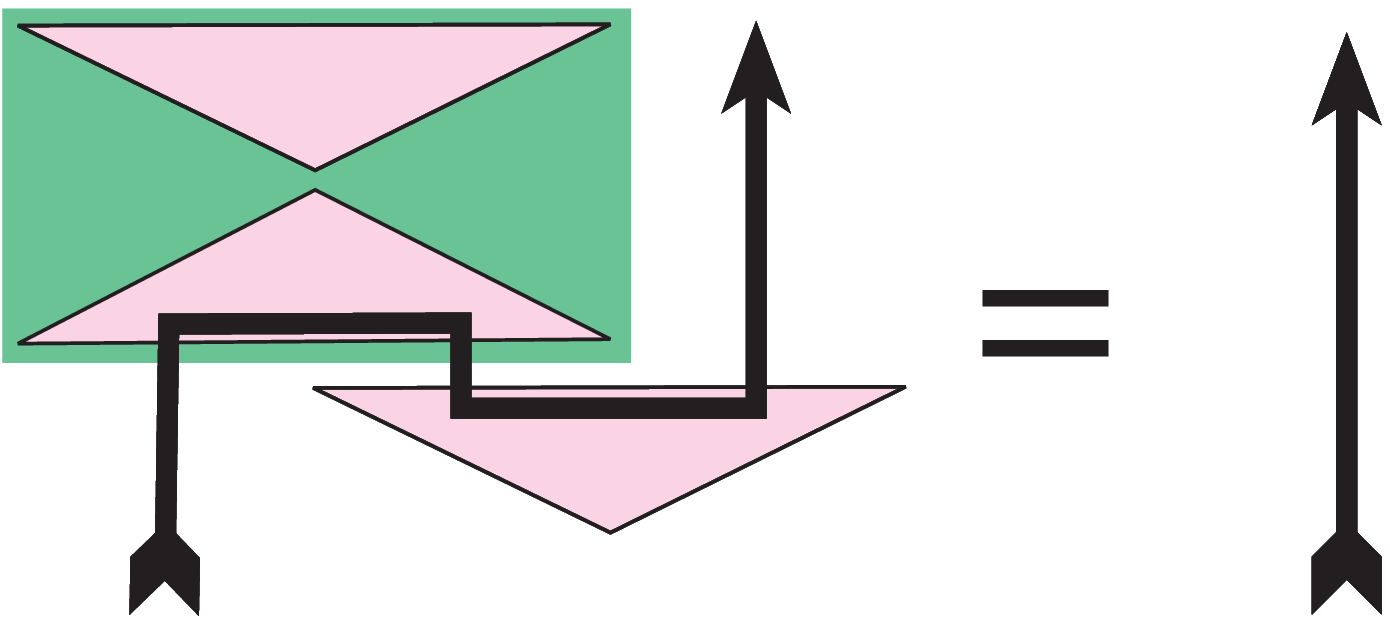,width=200pt}}     
\end{minipage}
\par\vspace{1mm}\noindent
However, as any physicist knows, we cannot impose a projector on a system with certainty, but only with some probability as a component in the spectral decomposition of some measurement.  Hence we will need a picture for each of the projectors ${\rm P}_{f_i}$ in this spectral decomposition.  For the sake of simplicity of the argument we are (temporarly) going to approximate an $f$-labeled bipartite projector by:
\par\vspace{1mm}\noindent
\begin{minipage}[b]{1\linewidth}
\centering{\epsfig{figure=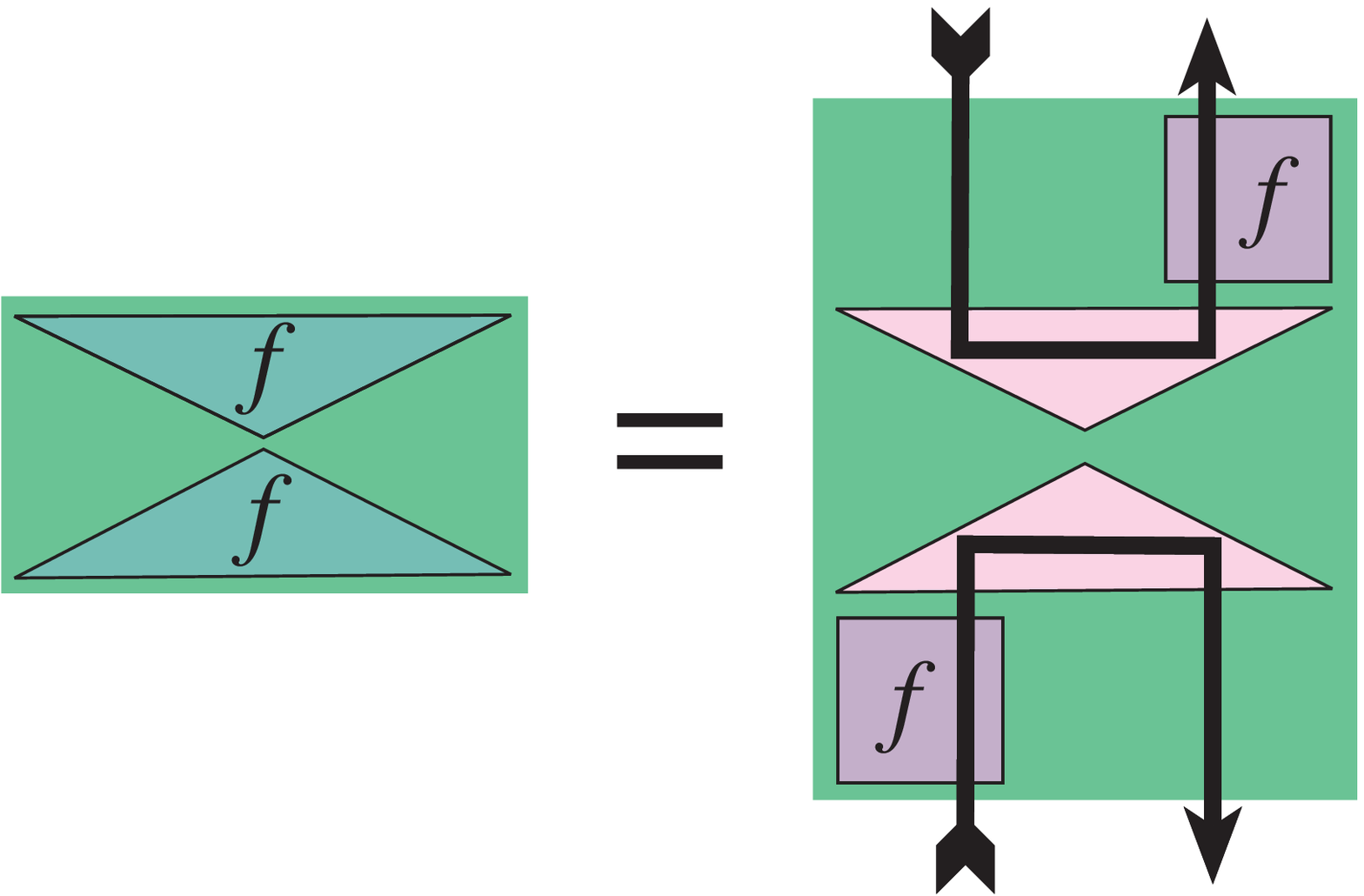,width=200pt}}     
\end{minipage}
\par\vspace{1mm}\noindent
which now has type ${\rm P}_f:A\otimes B^*\to A^*\otimes B$. Further down we will make up for this inaccuracy --- as we will show our approximation  corresponds to ignoring a pair of complex conjugations which in the standard teleporatation protocol do not play any role anyway cf.~\cite{Coe}.  Using the compositionally lemma {\bf(4)} and  introducing the required unitary correction $f_i^{-1}$ we obtain a \em full description of the teleportation protocol including correctness proof \em:\footnote{Compare this to the description and proof one finds in quantum computing textbooks and which only applies to the case of teleporting a qubit:
%%%%%%%%%%%%%%%%%%%%%%%%%%%
\par{\tiny 
\begin{center}
\fbox{ \begin{minipage}[b]{7.5cm} {\bf Description.} Alice has an `unknown' qubit $|\phi\rangle$ and wants to  send it to Bob.  They have the ability to communicate classical bits, and they share an entangled pair in the EPR-state, that is  ${1\over\sqrt{2}}(|00\rangle+|11\rangle)$, which Alice produced by first applying a {\sf Hadamard}-gate ${1\over\sqrt{2}}\left(\begin{array}{rr} 
1&1\\ 1&\!\!\!\!-1 \end{array}\right)$ to the first qubit of a qubit pair in the ground state $|00\rangle$, and by then applying a {\sf CNOT}-gate, that is 
\[ \left|00\right\rangle \mapsto \left|00\right\rangle  \ \ \   \left|01\right\rangle  \mapsto
\left|01\right\rangle  \ \ \   \left|10\right\rangle \mapsto  \left|11\right\rangle 
\ \ \   \left|11\right\rangle \mapsto \left|10\right\rangle\!, \] then {\bf
she sends the first qubit of the pair to Bob}.      To teleport her qubit, Alice  first performs a bipartite measurement on the unknown qubit and her half of the entangled pair in the Bell-base, that is \[ \left\{|0x\rangle+ (-1)^{z}|1(1-x)\rangle\bigm| x,z\in\{0,1\}\right\}\!, \]
where we denote the four possible outcomes of the measurement by $xz$.  Then {\bf she sends the 2-bit outcome $xz$ to Bob using the classical channel}. Then,  if $x=1$, Bob performs the unitary operation $\sigma_x=  \left(\begin{array}{rr}  0&1\\ 1&0 \end{array}\right)$ on its half of the shared entangled pair, and he also performs a unitary operation $\sigma_z=  \left(\begin{array}{rr} 1&\!\!\!\!0\,\\   0&\!\!\!\!-1 \end{array}\right)$ on it if $z=1$. Now {\bf Bob's half of the initially entangled pair is in state  $|\phi\rangle$}. \end{minipage}}\hspace{8mm}\fbox{
\begin{minipage}[b]{7.5cm}\noindent{\bf Proof.}  In the case that the measurement outcome of the Bell-base measurement is $xz$, for  \[ {\rm P}_{xz}:=\langle 0x+(-1)^{z} 1(1-x)|-\rangle|0x+ (-1)^{z}1(1-x)\rangle \] we have to apply ${\rm P}_{xz}\otimes {\sf id}$ to the input  state $|\phi\rangle\otimes{1\over\sqrt{2}}(|00\rangle+|11\rangle)$. Setting $|\phi\rangle =\phi_0|0\rangle+\phi_1|1\rangle$ we rewrite the input as \[  {1\over\sqrt{2}}(\phi_0|000\rangle+\phi_0|
011\rangle+\phi_1|100\rangle+\phi_1|111\rangle)={1\over\sqrt{2}}(\phi_0\!\!\sum_{x=0,1}\!\!|0xx\rangle+\phi_1\!\!\sum_{x=0,1}\!\!|1(1-x)(1-x)\rangle)
\] and application of ${\rm P}_{xz}\otimes {\sf id}$ then yields
${1\over\sqrt{2}}|0x+(-1)^{z}1(1-x)\rangle\otimes
(\phi_0|x\rangle+(-1)^{z}\phi_1|1-x\rangle)$. {\bf There are four cases concerning the unitary corrections $U_{xz}$ which have to be applied}.  For $x=z=0$ the third qubit is
$\phi_0|0\rangle+\phi_1|1\rangle=|\phi\rangle$. {\bf If $x=0$ and
$z=1$} it is $\phi_0|0\rangle-\phi_1|1\rangle$ which after applying $\sigma_z= \left(\begin{array}{rr} 1&\!\!\!\!0\,\\   0&\!\!\!\!-1 \end{array}\right)$ becomes $|\phi\rangle$. {\bf If} $x=1$ it is $\phi_0|1\rangle+(-1)^{z}\phi_1|0\rangle$ which after applying $\sigma_x= \left(\begin{array}{rr}  0&1\\ 1&0\end{array}\right)$ brings us back to the previous two cases, what completes this proof.\hfill$\Box$\end{minipage}}\end{center}\par}
%%%%%%%%%%%%%%%%%%%%%%%%%%%
\par\noindent 
This very `informal' discussion is far less intuitive and memorable, in other words, there's no `logic' to it.}
\par\vspace{3mm}\noindent
\begin{minipage}[b]{1\linewidth}
\centering{\fbox{\qquad\epsfig{figure=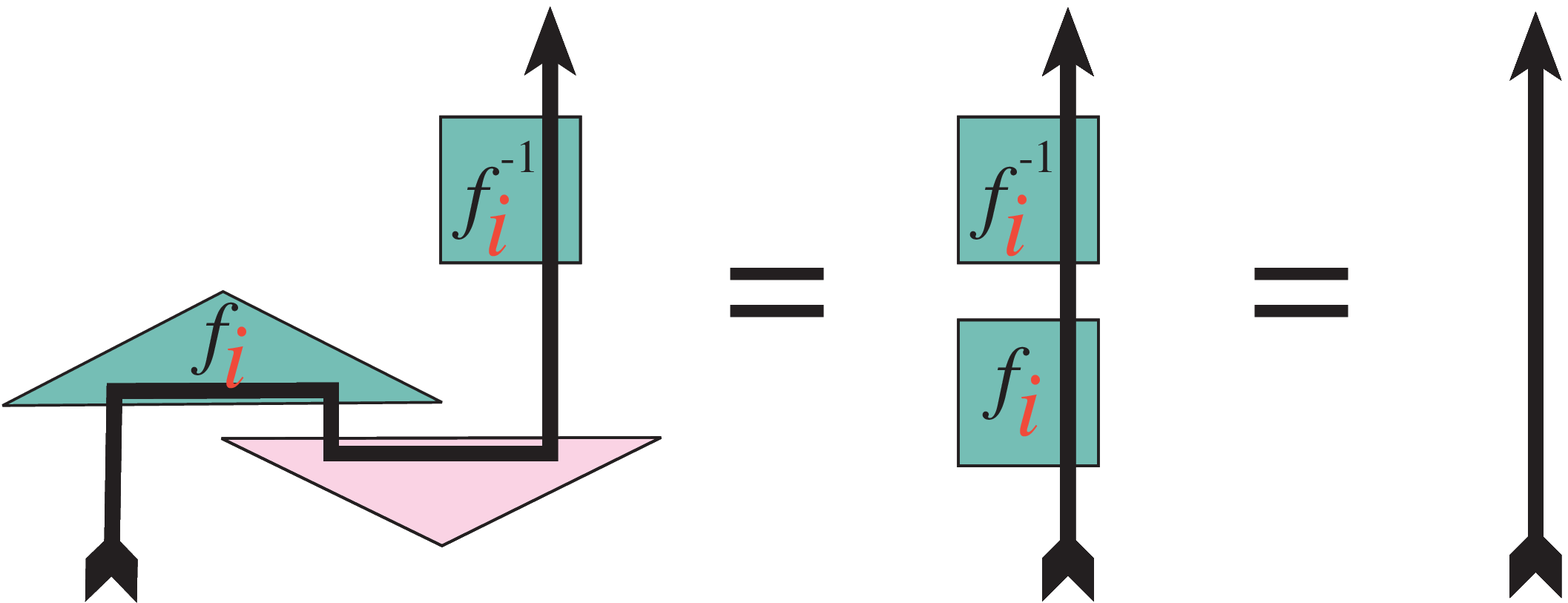,width=270pt}\qquad\hfill{\bf (7)}}}     
\end{minipage}
\par\vspace{3mm}\noindent
The classical communication is now implicit in the fact that the index $i$ is both present in the costate (= measurement-branch) and the correction, and hence needs to be send from Alice to Bob. Analogously, logic-gate teleportation \cite{Gottesman} (i.e.~teleporting while at the same time subjecting the teleported state to some `virtual' operation) arises as:
\par\vspace{1mm}\noindent
\begin{minipage}[b]{1\linewidth}
\centering{\epsfig{figure=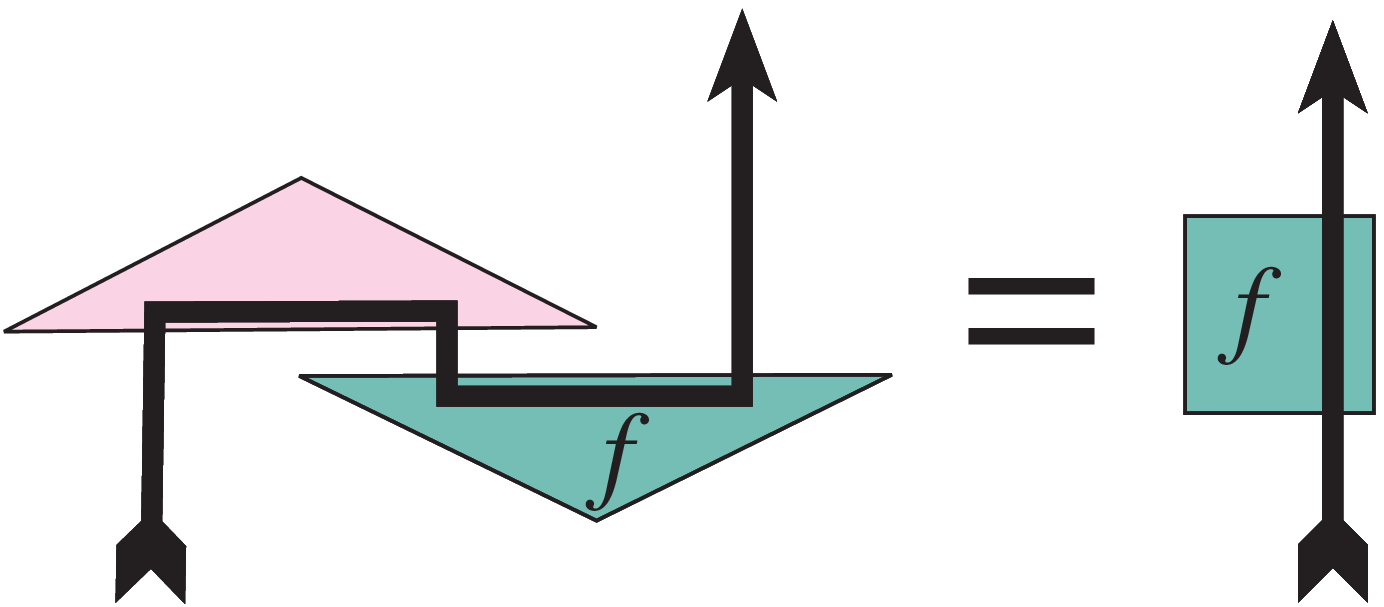,width=180pt}}     
\end{minipage}
\par\vspace{1mm}\noindent
where $f$ is the imposed operation, and entanglement swapping \cite{Swap} arises as: 
\par\vspace{1mm}\noindent
\begin{minipage}[b]{1\linewidth}
\centering{\epsfig{figure=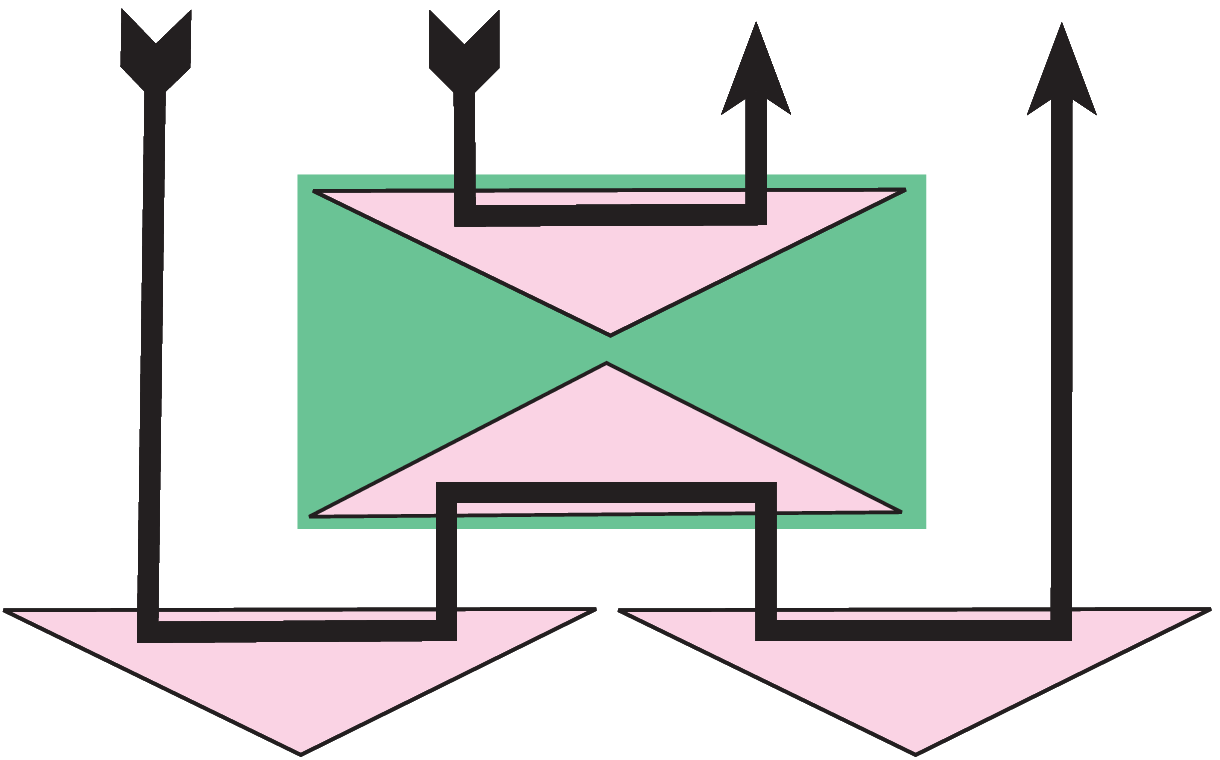,width=150pt}}     
\end{minipage}
\par\vspace{1mm}\noindent
One easily comes up with all kinds of variants of this scheme, for which we refer the reader to \cite{Coe}.
%\subsection{3.d. The Logic of Entanglement}

\subsection{3.d. The Hilbert space Model}

(i) We take \epsfig{figure=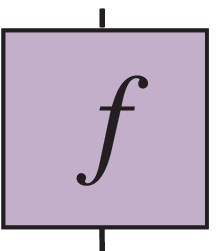,width=12pt} to be a linear map $f:{\cal H}_1\to{\cal H}_2$, 
we take \epsfig{figure=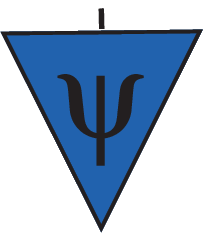,width=12pt} to be a linear map 
$\psi:\mathbb{C}\to{\cal H}$ which by linearity is completely determined by the vector $\psi(1)\in{\cal H}$, and we take \epsfig{figure=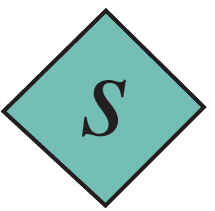,width=12pt}
to be the linear map $s:\mathbb{C}\to\mathbb{C}$ which is completely determined by the number $s(1)\in\mathbb{C}$.  (ii) We take parallel composition to be the tensor product and sequential composition to be ordinary composition of functions. (iii) We take ${\cal H}^*$ to be the conjugate Hilbert space of ${\cal H}$, that is, the Hilbert space with the same elements as ${\cal H}$ but with inner-product  and scalar multiplication conjugated (see \cite{AC1,AC1.5} for details), and we take $f^\dagger$ to be the (linear) adjoint of $f$.  
(iv) we take \epsfig{figure=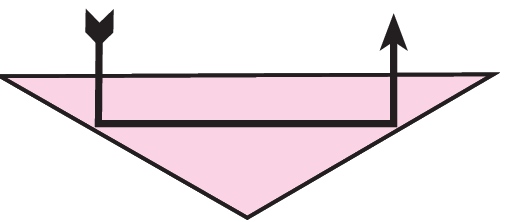,width=25pt} to be the linear map
\[
\mathbb{C}\to{\cal H}^*\otimes{\cal H}::1\mapsto\bigm|\sum_i e_i\!\otimes\! e_i\bigr\rangle
\]
and hence, the corresponding costate \epsfig{figure=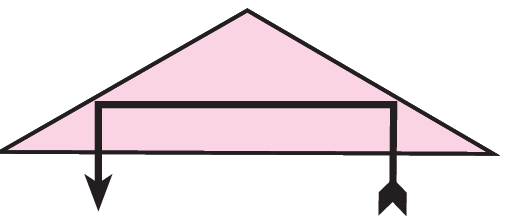,width=25pt} is its adjoint
\[
{\cal H}^*\otimes{\cal H}\to\mathbb{C}::\Phi\mapsto\bigl\langle\sum_i e_i\!\otimes\! e_i\bigm|\Phi\bigr\rangle
\]
which is uqual to the linear extension of $\phi_1\!\otimes\phi_2\mapsto\langle\phi_1\mid\phi_2\rangle$. We can now verify the axiom.  Since we have that 
$\epsfig{figure=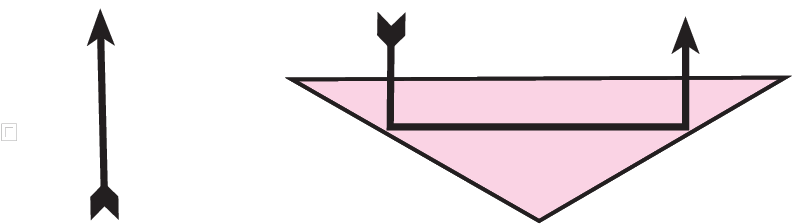,width=45pt}=(-)\otimes\bigl(\sum_i e_i\!\otimes\! e_i\bigr)
=\sum_i(- \otimes\!e_i)\!\otimes\! e_i$ it follows that
$\epsfig{figure=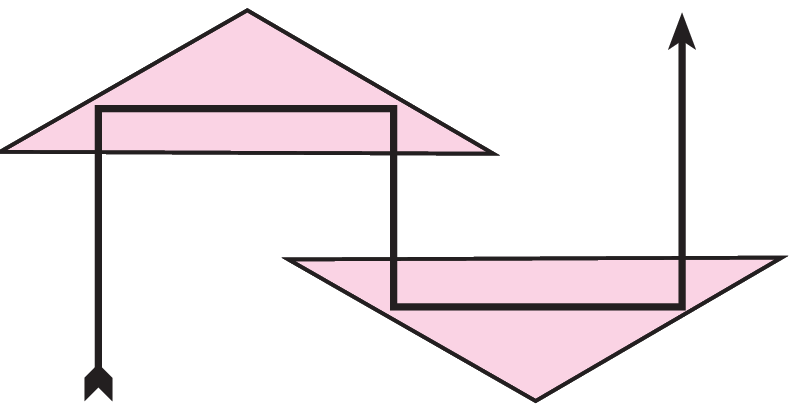,width=45pt}=\sum_i \langle - \mid\!e_i\rangle\cdot e_i =1_{\cal H}$
and that's it!  

\bigskip\par\hrule\bigskip

\centerline{{\bf Result:} \em All derivations in the picture calculus constitute a statement about quantum mechanics.\em}   

\bigskip\par\hrule\bigskip

We leave it as an easy but interesting exercise for the reader to show that each non-degenerate bipartite projector indeed decomposes as in picture {\bf(6)}. Its normalization translates in pictures as:
\par\vspace{3mm}\noindent
\begin{minipage}[b]{1\linewidth}
\centering{\fbox{\qquad\epsfig{figure=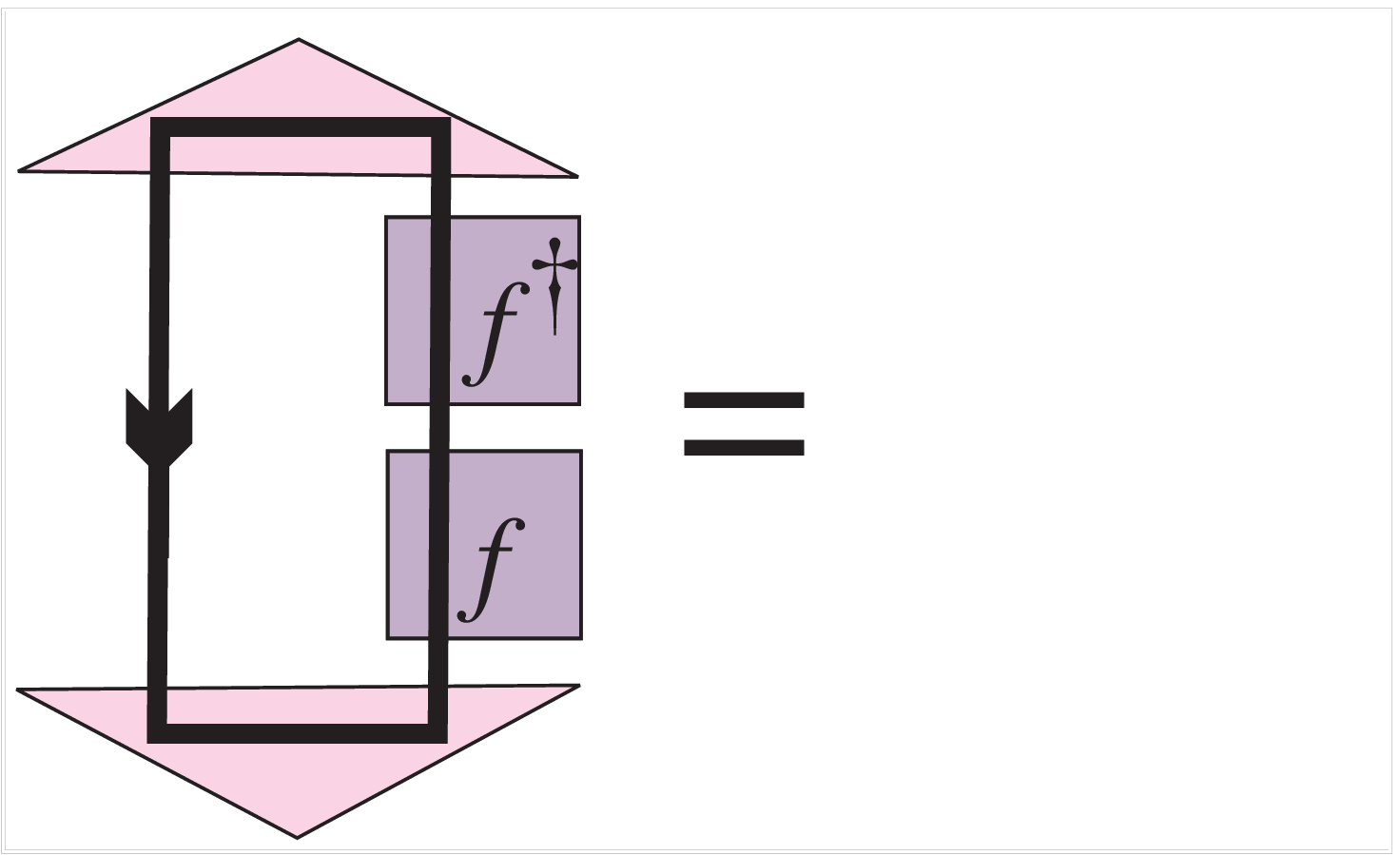,width=160pt}\ \ \hfill{\bf (8)}}}     
\end{minipage}
\par\vspace{3mm}\noindent
The void at the right side of the equalit is the identity diamond which has no input/output nor weight and hence is graphically completely redundant ($1\in\mathbb{C}$ in the Hilbert space case). In particular is this bipartite projector just a special case of a general \em normalized non-degenerate projector \em which depicts as:
\par\vspace{3mm}\noindent
\begin{minipage}[b]{1\linewidth}\centering{\fbox{\qquad
\epsfig{figure=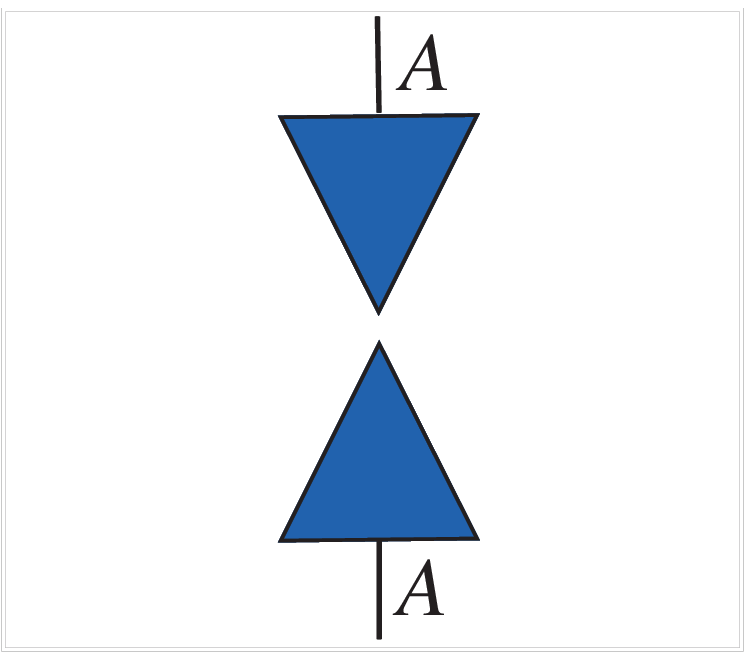,width=75pt}\qquad \qquad \qquad\epsfig{figure=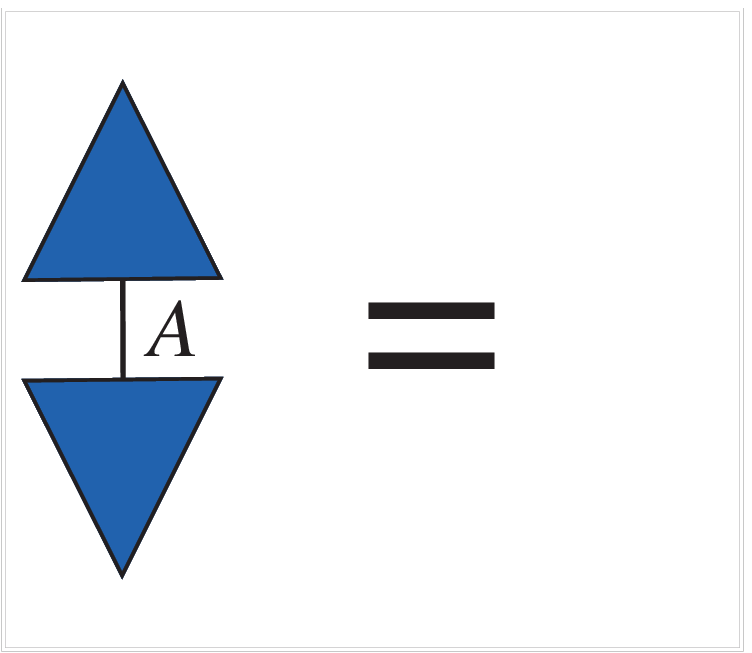,width=75pt}
\qquad\hfill{\bf (9)}}}\end{minipage}
\par\vspace{3mm}\noindent
where the equality with void righthandside provides idempotence:
\par\vspace{1mm}\noindent
\begin{minipage}[b]{1\linewidth}\centering{
\epsfig{figure=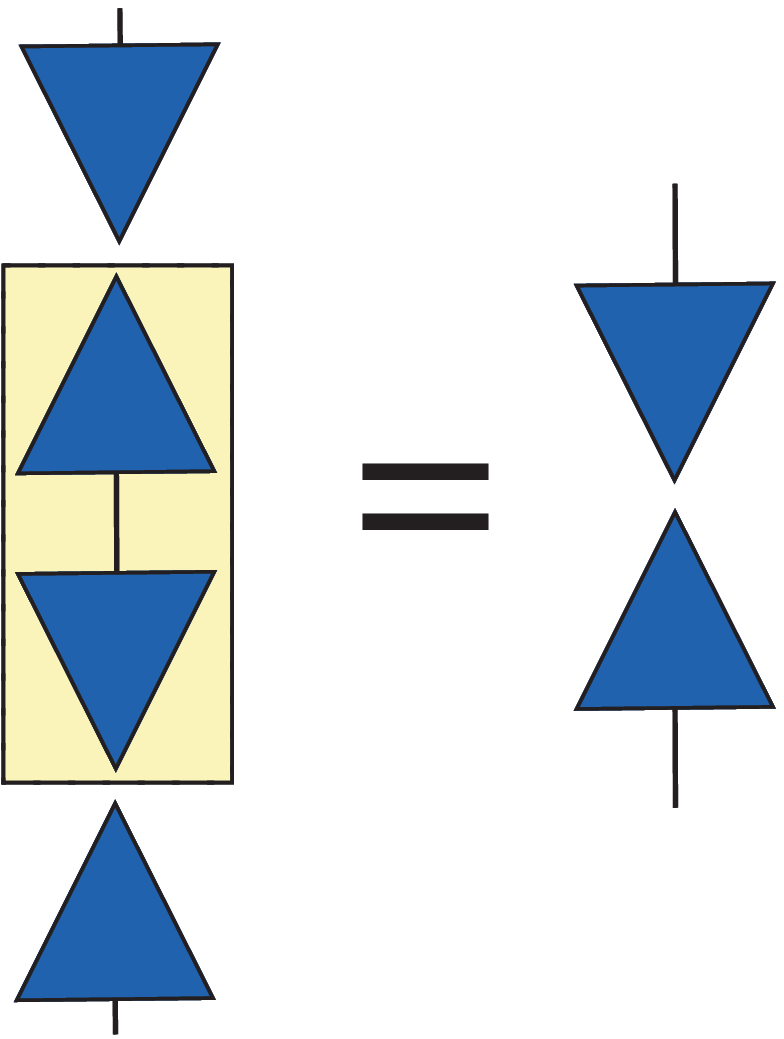,width=75pt}
}\end{minipage}
\par\vspace{0mm}\noindent
Such a normalized non-degenerate projector is on its turn a special case of a general \em normalized projector\em:
\par\vspace{3mm}\noindent
\begin{minipage}[b]{1\linewidth}\centering{\fbox{\qquad
\epsfig{figure=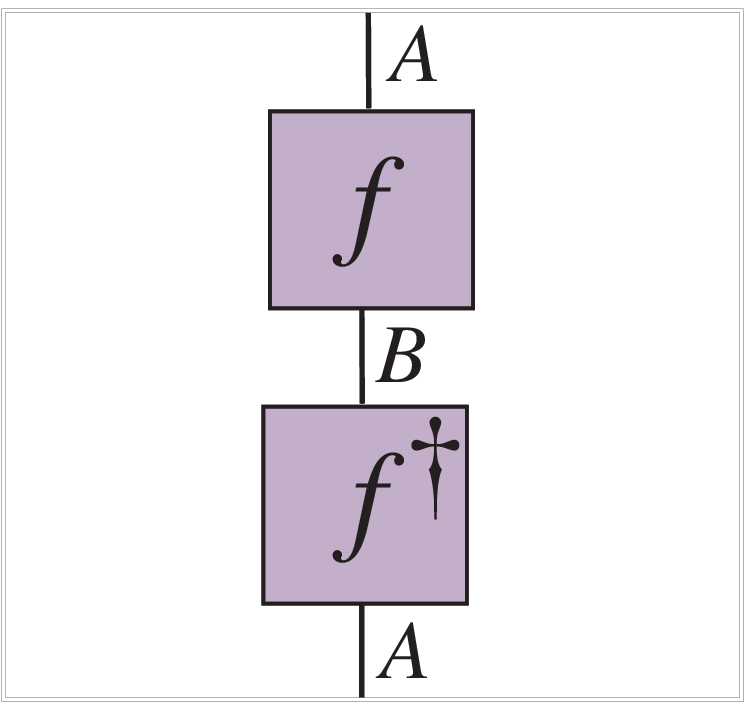,width=75pt}\qquad \qquad \qquad\epsfig{figure=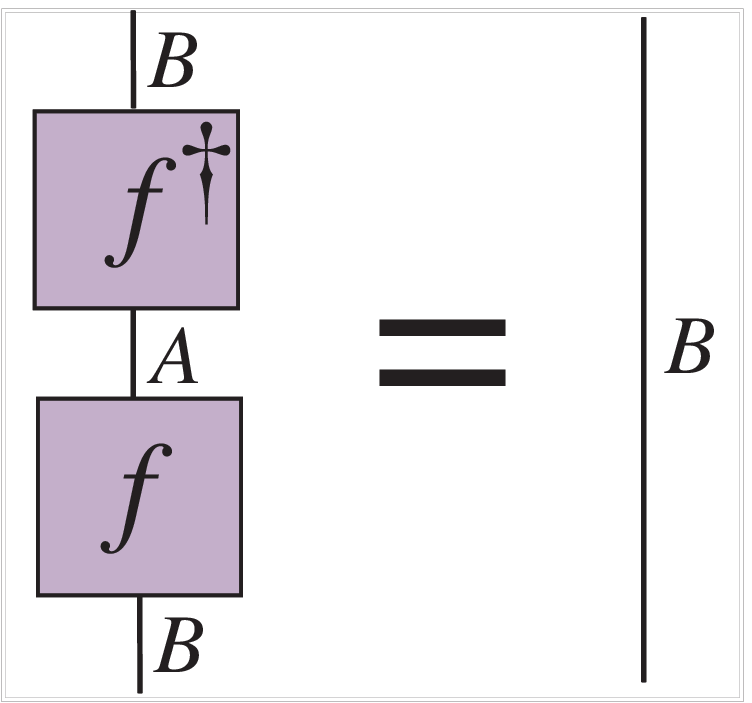,width=75pt}
\qquad$\!\!\!$\hfill{\bf (10)}}}\end{minipage}
\par\vspace{3mm}\noindent
Note also that the shape $f\circ f^\dagger$ expresses \em positivity \em at the level of the picture calculus.

The reason that diamonds, states and costates \em are allowed to behave non-local in the plane of the picture \em is due to the fact that $\mathbb{C}$ is a \em natural unit for the tensor\em, which means that there are isomorphisms $\lambda_{\cal H}:{\cal H}\to\mathbb{C}\otimes{\cal H}$ and $\rho_{\cal H}:{\cal H}\to{\cal H}\otimes \mathbb{C}$ such that the following two diagrams commute for all $f: {\cal H}_1\to{\cal H}_2$:
\begin{diagram}
{\cal H}_1&\rTo^f&{\cal H}_2&&&{\cal H}_1&\rTo^f&{\cal H}_2\\
\dTo^{\lambda_{{\cal H}_1}}& &\dTo_{\lambda_{{\cal H}_2}}
&&&
\dTo^{\rho_{{\cal H}_1}}& &\dTo_{\rho_{{\cal H}_2}}\\
\mathbb{C}\otimes{\cal H}_1&\rTo^{1_{\mathbb{C}}\otimes f}&\mathbb{C}\otimes{\cal H}_2 &&&
{\cal H}_1\otimes \mathbb{C}&\rTo^{f\otimes 1_{\mathbb{C}}}&{\cal H}_2\otimes \mathbb{C}
\end{diagram}
One can verify that all the nice properties scalar multiplication admits are due to this fact and this fact only \cite{AC1,AC1.5, KellyLaplaza}.

\subsection{3.e. Dirac's notation}

Dirac's notation \cite{Dirac} is a small instance of what is going on here and one could wonder what the quantum formalism would have been by now in the case Dirac had the right kind of category theory at his possession. Since we have:
\par\vspace{-1mm}\par\noindent
\begin{center}
\begin{picture}(360,40)
\put(42,12){$\bR\mid\e f\circ\psi\bR\rangle\e=$}
\put(92,-7){\epsfig{figure=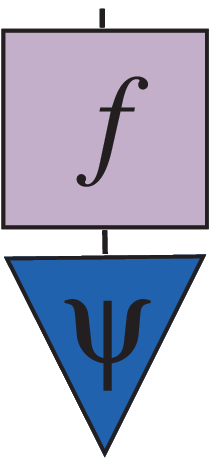,width=20pt}}
%\put(122,12){$=\, f\circ\psi$}
\put(234,12){$\bR\langle\e f\circ\phi\bR\mid\e\ =$}
\put(285,-7){\epsfig{figure=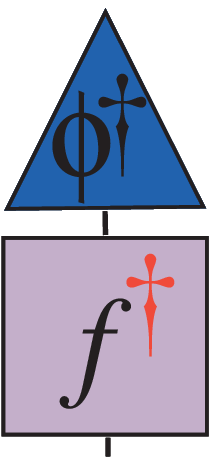,width=20pt}}
%\put(315,12){$=\phi^\dagger\circ f^\dagger$}
\end{picture}
\end{center}\par\vspace{-1mm}\par\noindent
adjointness yields its defining property through pictures:
\par\vspace{1mm}\noindent
\begin{center}
\begin{picture}(140,50)
\put(0,22){$\bR\langle\e f\circ\phi\bR\mid\e\psi\bR\rangle\e=$}
\put(62,-07){\epsfig{figure=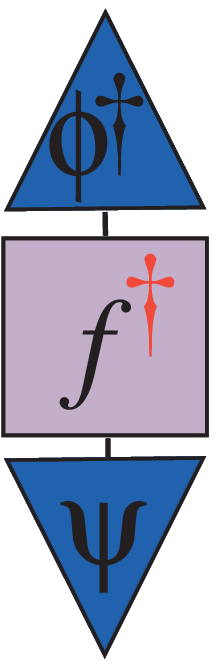,width=20pt}}
\put(92,22){$=\bR\langle\e\phi\bR\mid\e f^{\bR\dagger\e}\circ\psi\bR\rangle\e$}
\end{picture}
\end{center}
\par\vspace{-1mm}\par\noindent
and since unitarity means:
\par\vspace{1mm}\noindent
\begin{minipage}[b]{1\linewidth}
\centering{\ \epsfig{figure=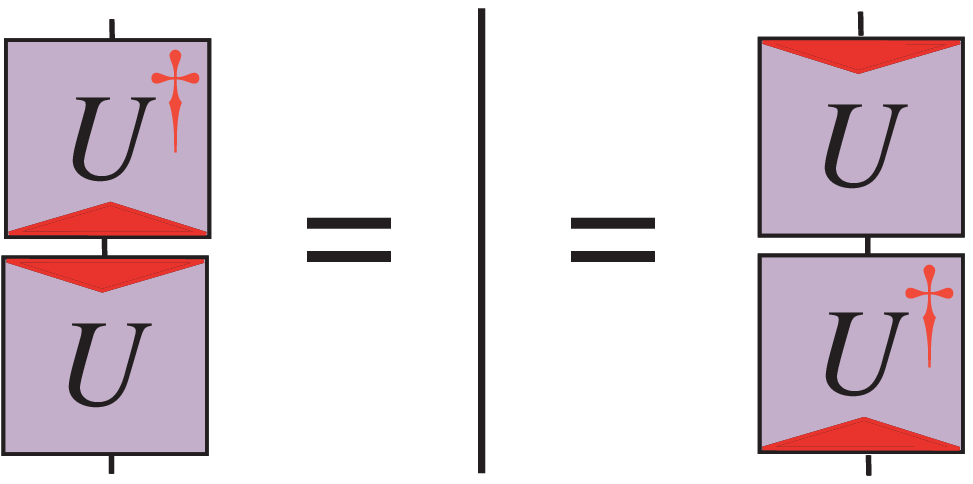,width=85pt}}     
\end{minipage}
\par\vspace{1mm}\noindent
we obtain preservation of the inner-product:
\par\vspace{1mm}\noindent
\begin{center}\begin{picture}(240,60)
\put(0,22){$\bR\langle\e U\circ \phi\bR\mid\e U\circ\psi\bR\rangle\e=$}
\put(83,-15){\epsfig{figure=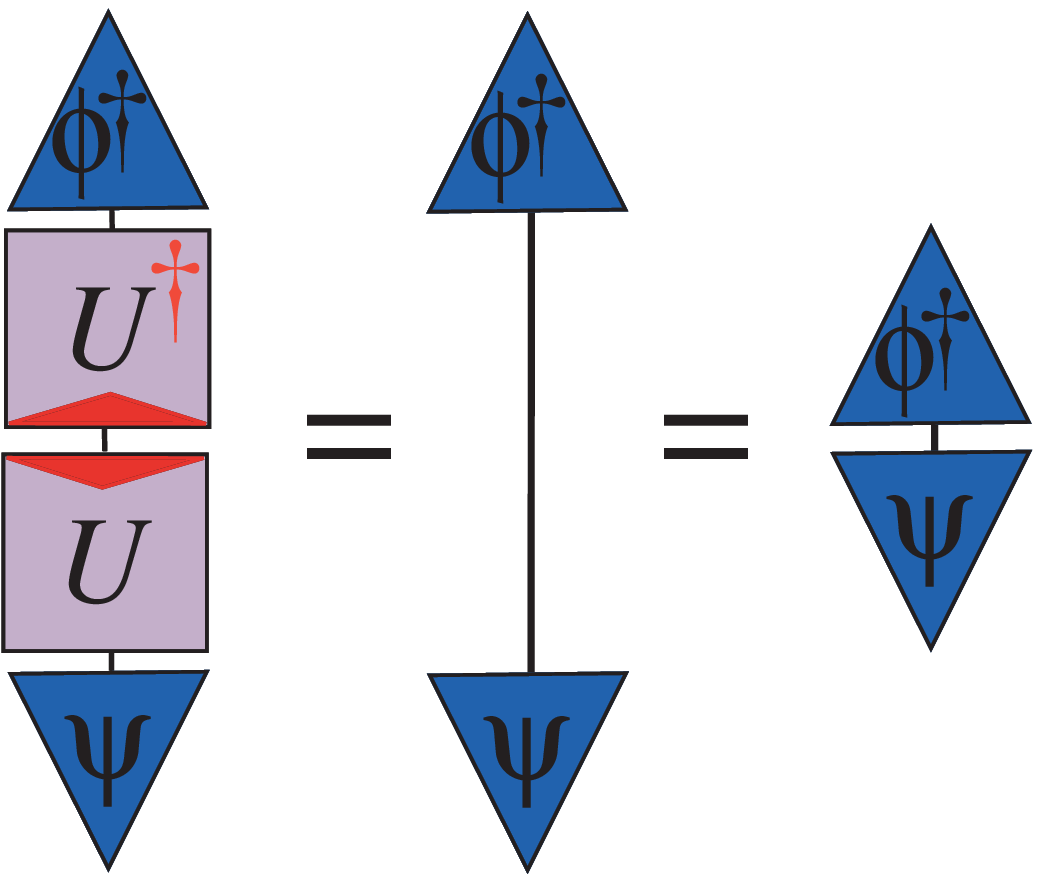,width=95pt}}
\put(188,22){$=\bR\langle\e \phi\bR\mid\e \psi\bR\rangle\e$}
\end{picture}
\end{center}
\par\vspace{-1mm}\par\noindent

\section{4. MORE theorem proving}

Now that we have  established the fact that Hilbert space quantum mechanics is a particular incarnation of the picture calculus we continue deriving results within it. These will in particular expose that the key structural ingredients of quantum mechnics such as complex conjugation, transposition, unitarity, inner-product, Hilbert-Schmidt norm and inner-product, trace, normalization, elimination of global phases, complete positivity (of which many might seem extremely closely related to the particular Hilbert space model) all do exist at the general level of the pictures.
%\subsection{4.a. Adjoints and Unitarity}

\subsection{4.a. Hilbert-Schmidt Correspondence}

In all the above a key role is implicitly played by the fact that linear maps of type ${\cal H}_1\!\to\!{\cal H}_2$ and the bipartite vectors ${\cal H}_1^{*}\otimes{\cal H}_2$ are in bijective correspondence. An easy way to see this is by considering the matrix of a linear map $f=\sum_{ij} m_{ij}| j\rangle\langle i|$ and observing that it determines the bipartite state $\Psi=\sum_{ij} m_{ij}
|ij\rangle$ in a bijective manner.  In fact, the truly \em natural \em isomorphism is:
\[
{\cal H}_1\!\to\!{\cal H}_2\ \simeq\ {\cal H}_1^{*}\otimes{\cal H}_2
\]
i.e.~we need to conjugate the first Hilbert space (although this gives an isomorphic copy).
In the above argument establishing the bijection we used the matrix representation of Hilbert spaces and hence essentially the whole vector space structure, but in fact we need none of this, and we will show that in \em any \em picture calculus we always have:
\par\vspace{3mm}\noindent
\begin{minipage}[b]{1\linewidth}  
\centering{\fbox{\qquad\epsfig{figure=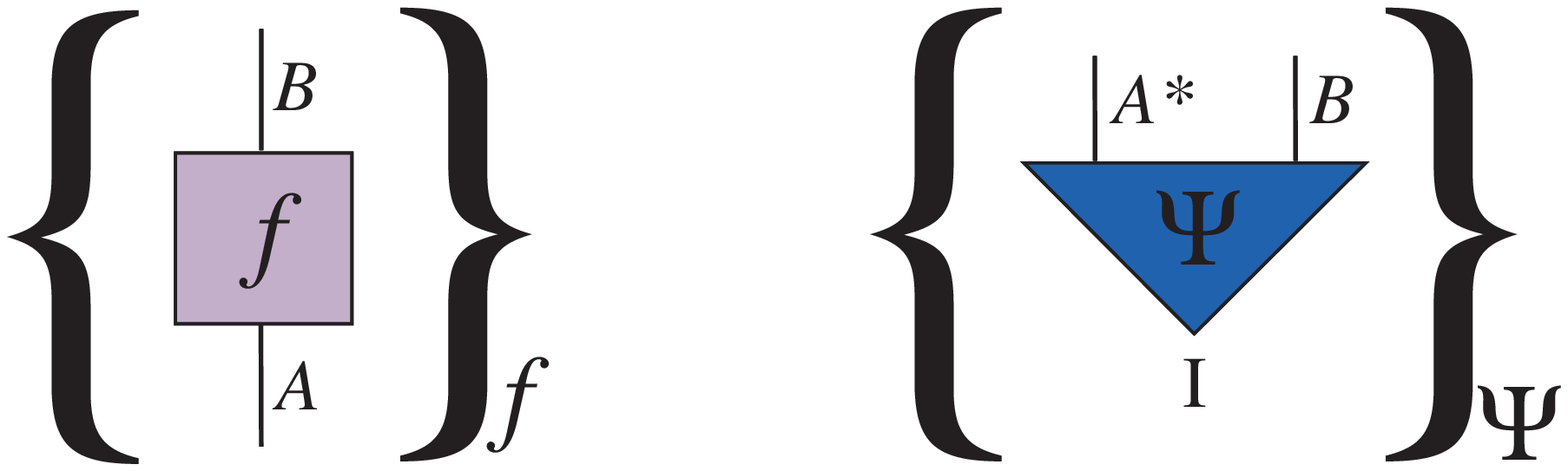,width=188pt}\qquad$\!\!\!$\hfill{\bf (11)}}}
\begin{picture}(340,0)
\put(152,38){$\simeq$}
\end{picture}
\end{minipage}
\par\vspace{0mm}\noindent
This is a weird statement, saying that there is a bijective correspondence between what at first sight seem to be independent primitive data nl.~boxes vs.~triangles. However it turns out that the assignment:
\par\vspace{2mm}\noindent
\begin{minipage}[b]{1\linewidth}  
\centering{\epsfig{figure=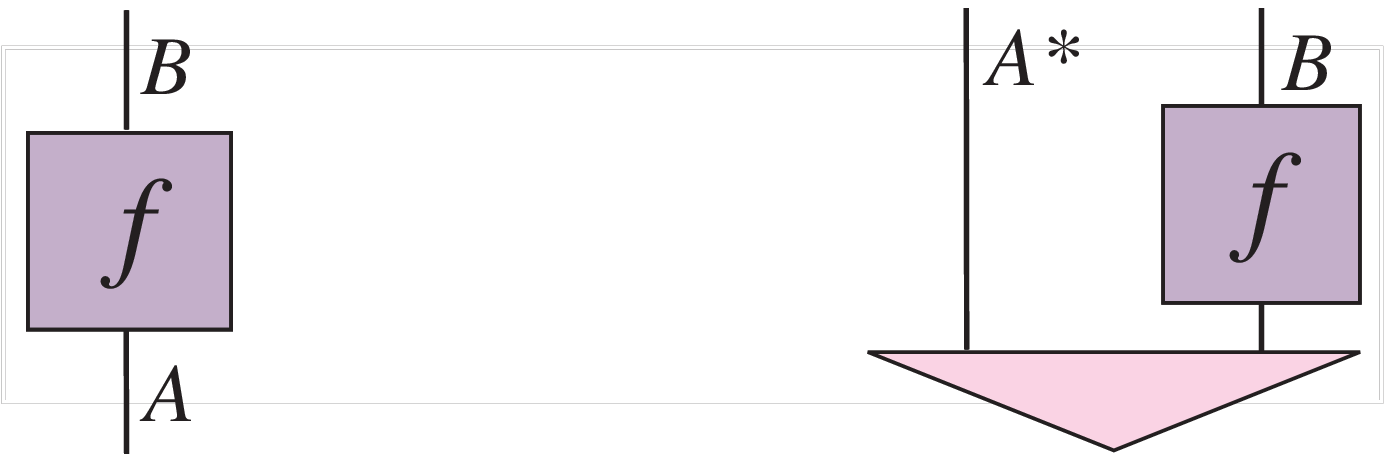,width=132pt}}
\begin{picture}(340,0)
\put(155,32){$\mapsto$}
\end{picture}
\end{minipage}
\par\vspace{-3mm}\noindent
defines a bijective correspondence due to the axiom.  First we prove injectivity of this assignment.
Given that 
\par\vspace{1mm}\noindent
\begin{center}
\epsfig{figure=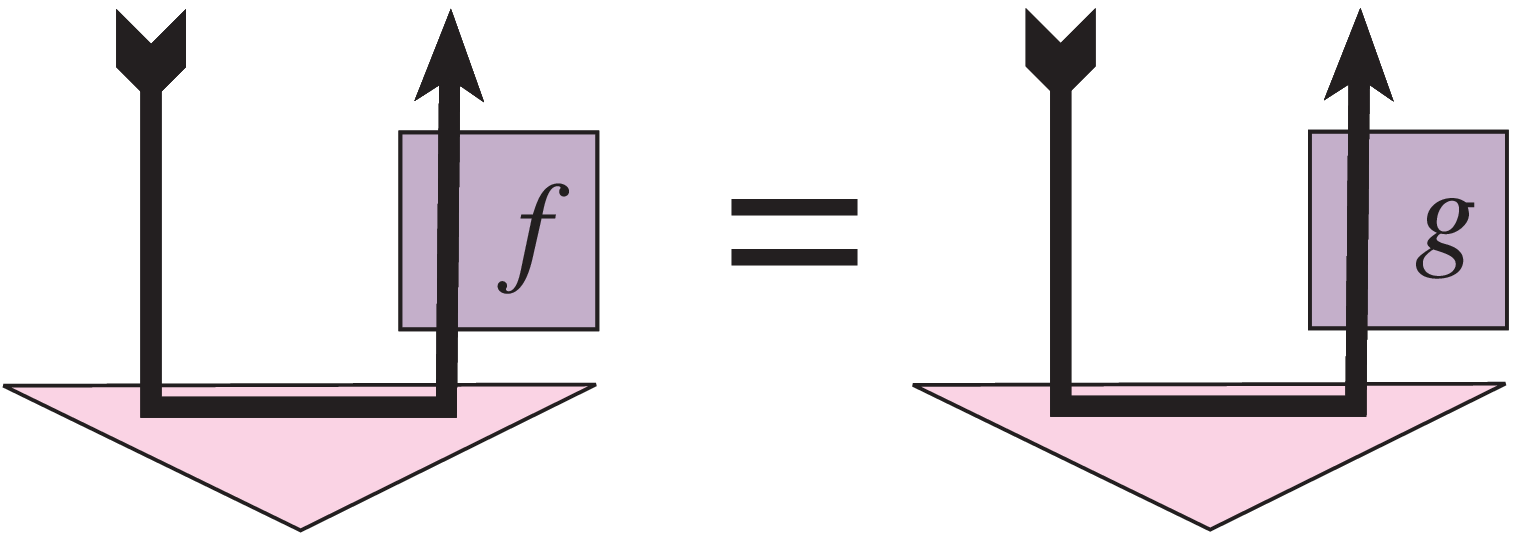,width=140pt}
\end{center}
\par\vspace{-1mm}\noindent
it follows by the axiom that:
\par\vspace{-1mm}\noindent
\begin{center}
\epsfig{figure=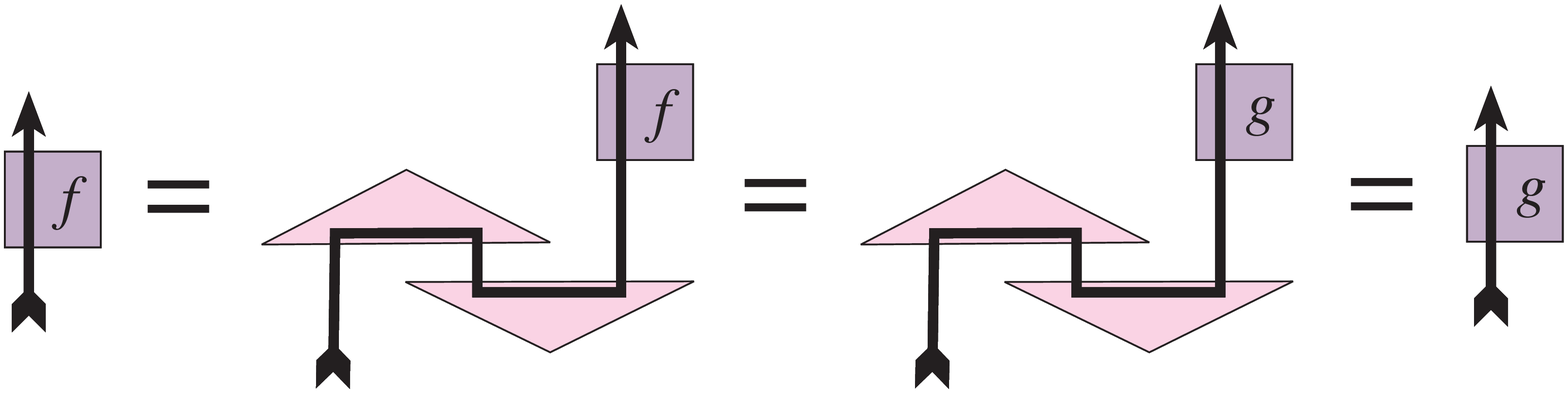,width=300pt}
\end{center}
\par\vspace{1mm}\noindent
Now we prove surjectivity i.e.~we need to find an operation which under the assignment yields any arbitrary bipartite state as its immage. Choosing the following operation as argument:
\par\vspace{1mm}\noindent
\begin{minipage}[b]{1\linewidth}  \centering{\epsfig{figure=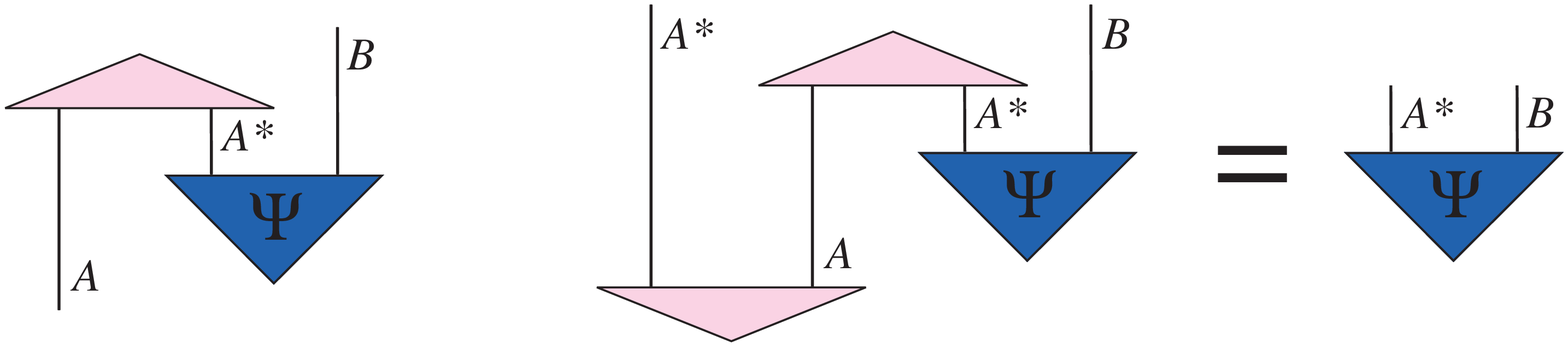,width=320pt}}
\begin{picture}(340,0)
\put(108,50){$\mapsto$}
\end{picture}
\end{minipage}
\par\vspace{-4mm}\noindent 
we indeed obtain any bipartite state as image, what completes the proof.
%Below we will denote the image of an operation $f$ under this bijection as $\uu f\uuu$. 

\subsection{4.b. Factoring the Adjoint}

We introduce two more derived notions for each operation:
\par\vspace{3mm}\noindent
\begin{minipage}[b]{1\linewidth}  
\centering{\fbox{\qquad\epsfig{figure=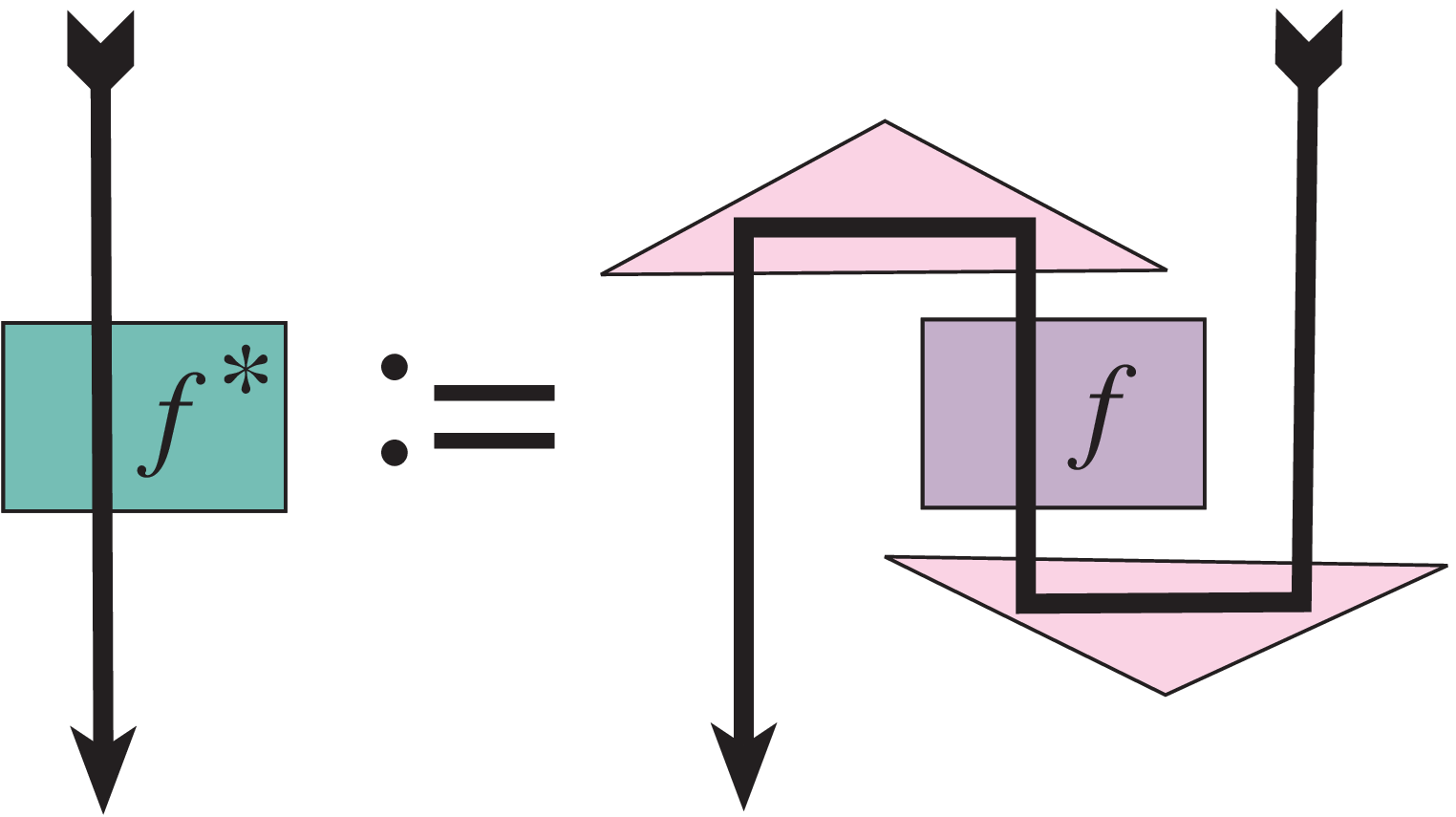,width=150pt}\qquad
\qquad\qquad\epsfig{figure=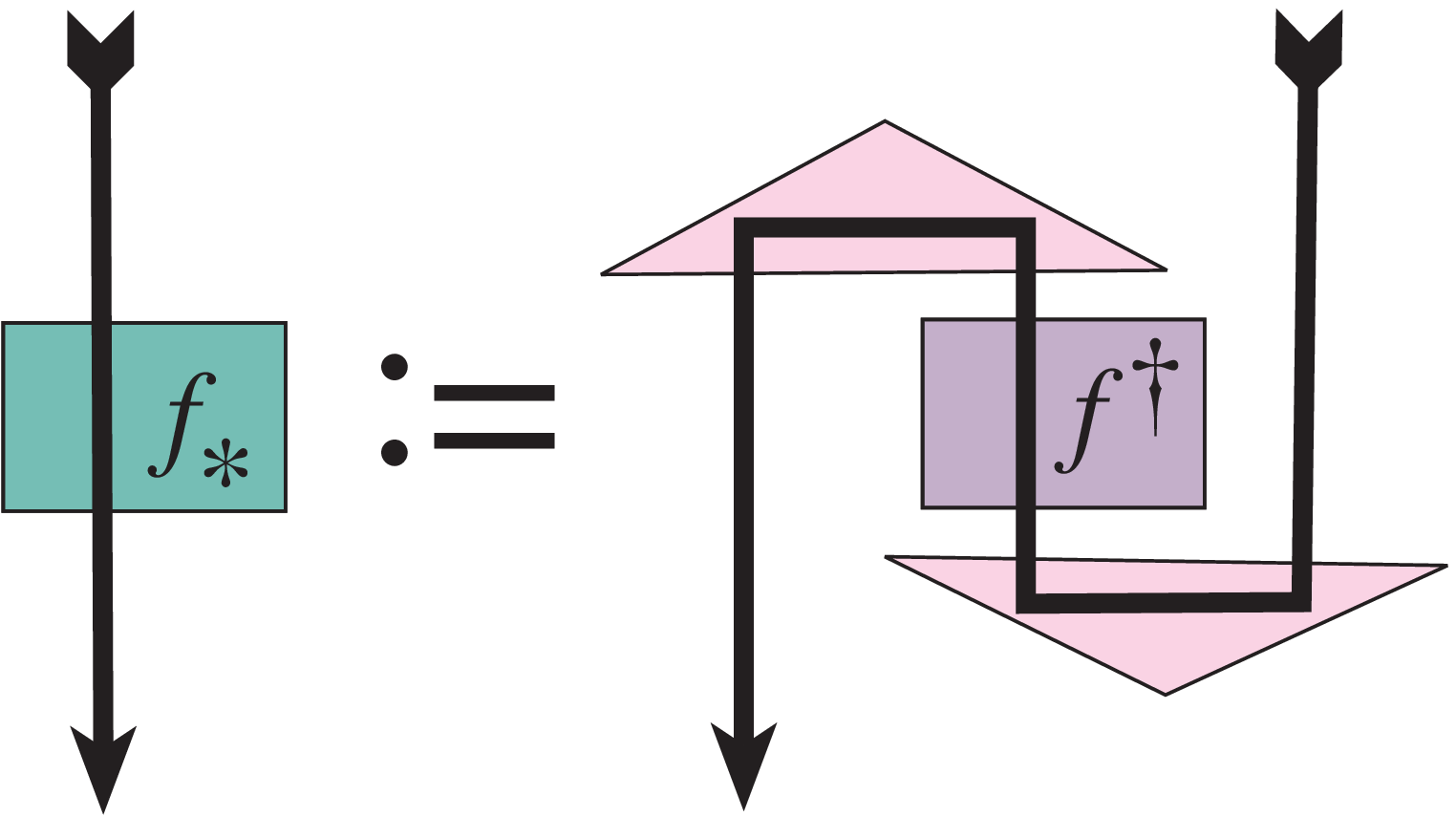,width=150pt}\qquad$\!\!\!$\hfill{\bf (12)}}}  
\end{minipage}
\par\vspace{3mm}\noindent
When we unfold these definitions we obtain:
\par\vspace{1mm}\noindent
\begin{minipage}[b]{1\linewidth}
\centering{\epsfig{figure=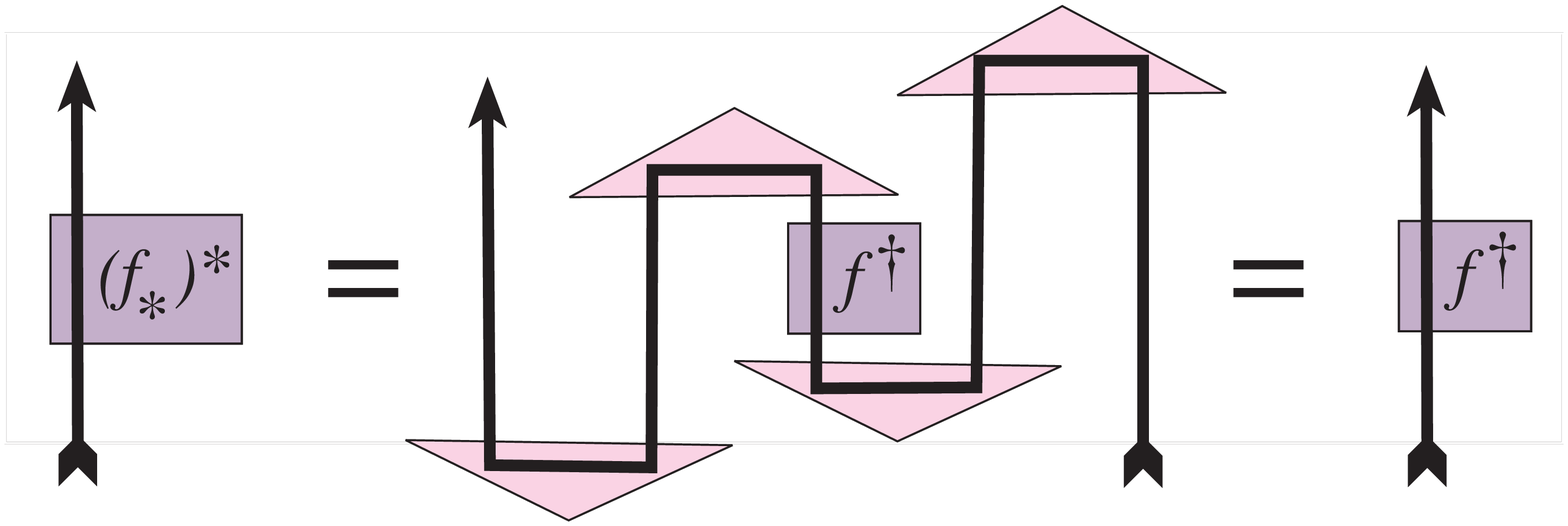,width=280pt}}
\end{minipage}
\par\vspace{1mm}\noindent 
and analogously one establishes that $((-)^*)_*=(-)^\dagger$. Hence we discovered that the adjoint $(-)^\dagger$ factors into the newly introduced operations $(-)^*$ and $(-)_*$. In particular do we have for the Hilbert space case that  
{\bf $(-)^*$ := transposition} and {\bf $(-)_*$ := complex conjugation},
and hence each picture calculus both has an analogue to transposition and an analogue complex conjugation, which combine into the adjoint.  

We now also have the right tools to revise the approximation of an $f$-labeled non-degenerate bipartite projector we used in Section {\bf 3.c} and correct it using picture {\bf (12)} into:
\par\vspace{1mm}\noindent
\begin{minipage}[b]{1\linewidth}
\centering{\epsfig{figure=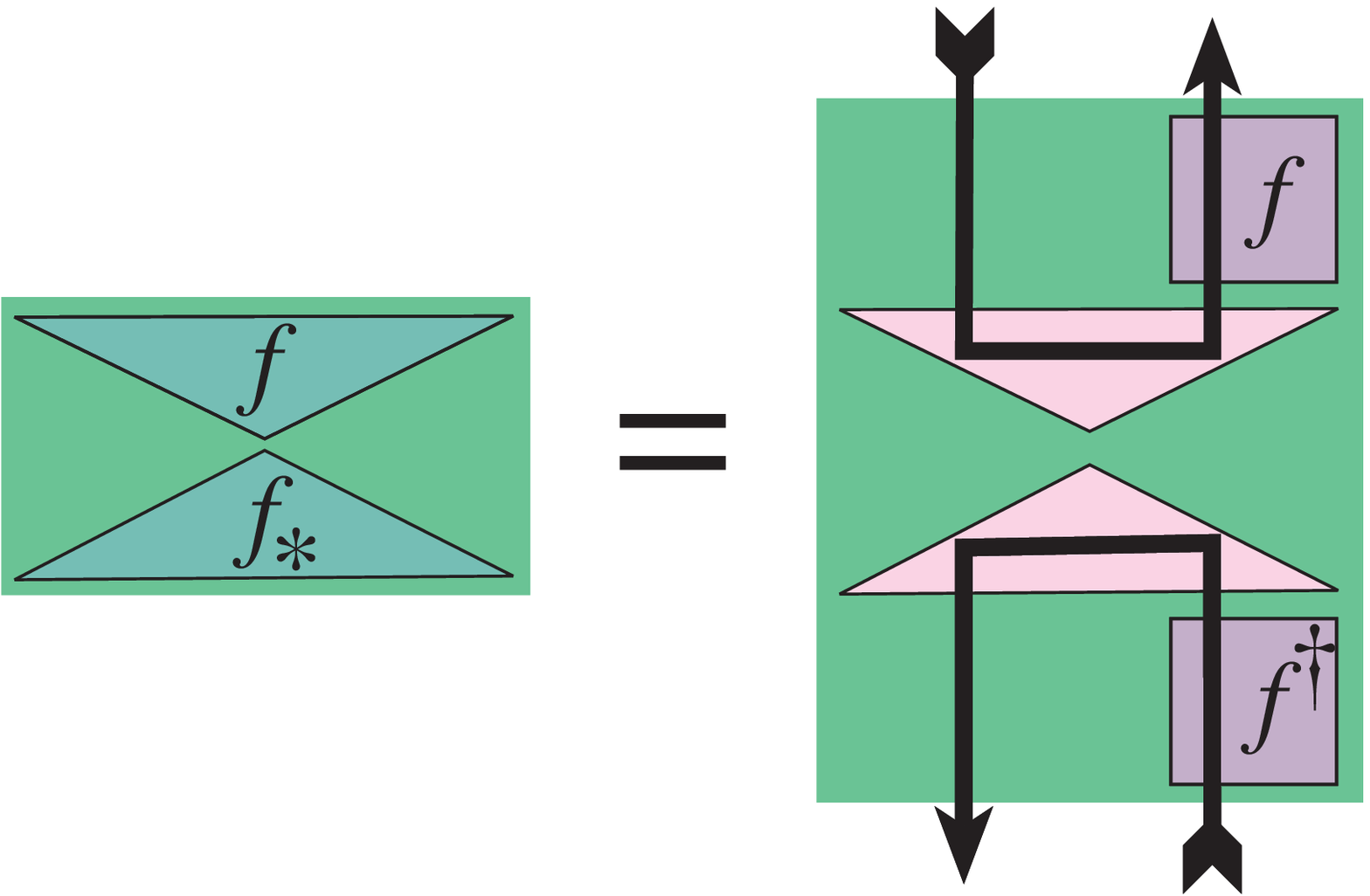,width=200pt}}     
\end{minipage}
%\par\vspace{1mm}\noindent

\subsection{4.c. Trace}

We define yet another operation called \em partial trace \em as:
\par\vspace{3mm}\noindent 
\begin{minipage}[b]{1\linewidth}  
\centering{\fbox{\qquad\epsfig{figure=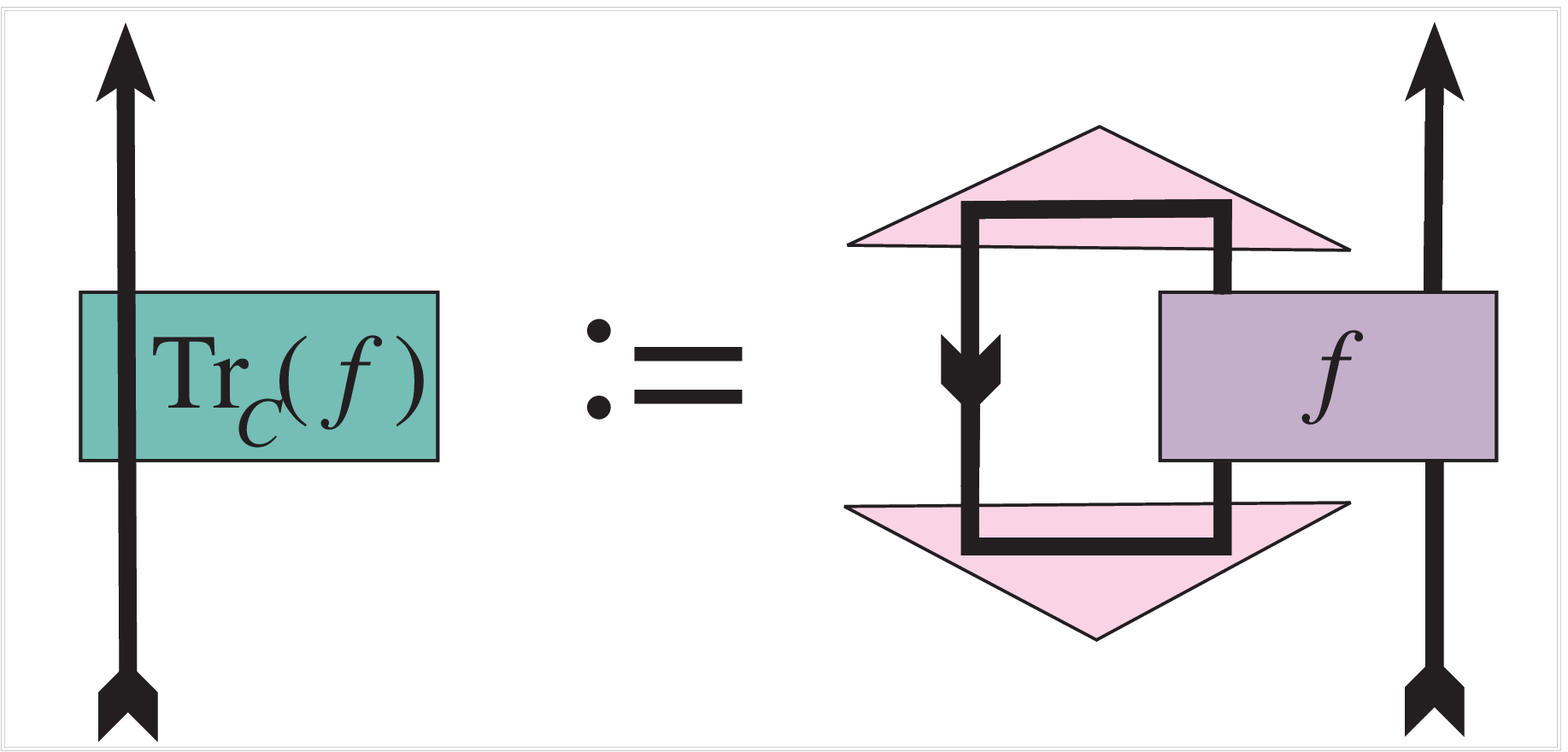,width=180pt}\qquad$\!\!\!$\hfill{\bf (13)}}}  
\end{minipage}
\par\vspace{3mm}\noindent 
which has the \em full trace \em as a particular case:
\par\vspace{3mm}\noindent 
\begin{minipage}[b]{1\linewidth}  
\centering{\fbox{\qquad\epsfig{figure=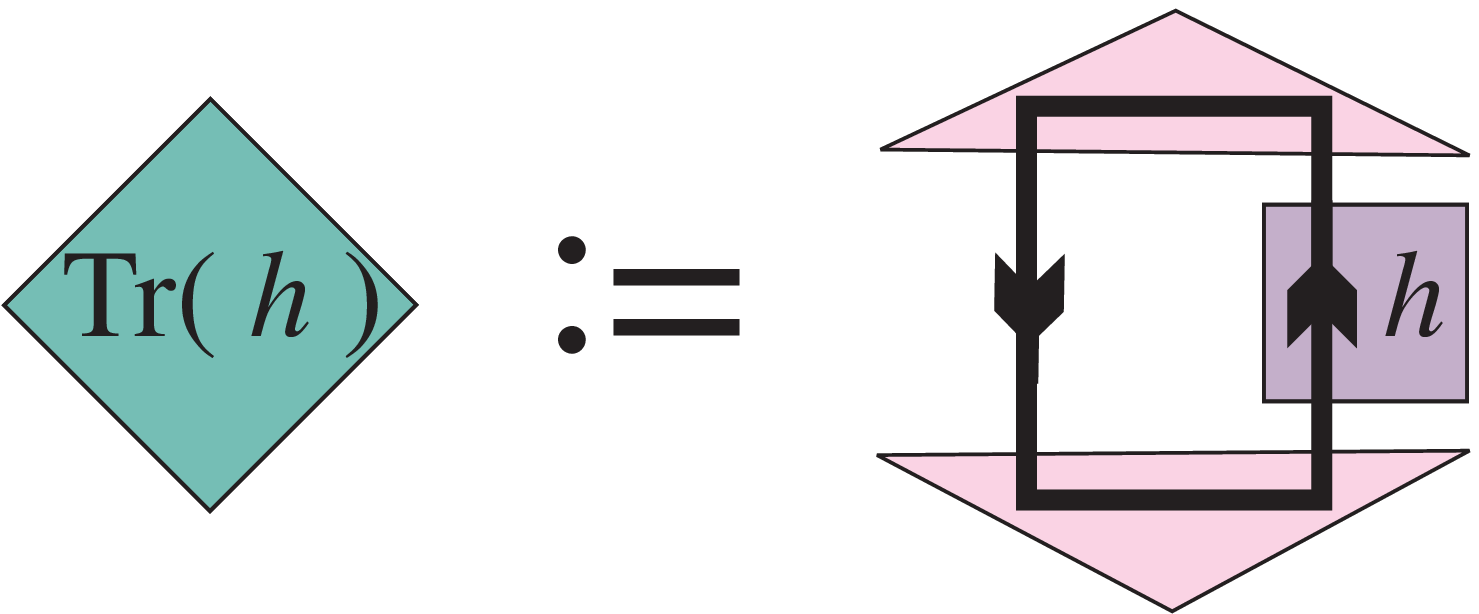,width=160pt}\qquad$\!\!\!$\hfill{\bf (14)}}}  
\end{minipage}
\par\vspace{3mm}\noindent 
This full trace assigns to each operation a diamond-shaped box which we can interpret as the \em probabilistic `weight' \em carried by this box.  We now prove a property of this full trace.  Again unfolding definitions we get:
\par\vspace{3mm}\noindent 
\begin{minipage}[b]{1\linewidth}  
\centering{\fbox{\qquad\epsfig{figure=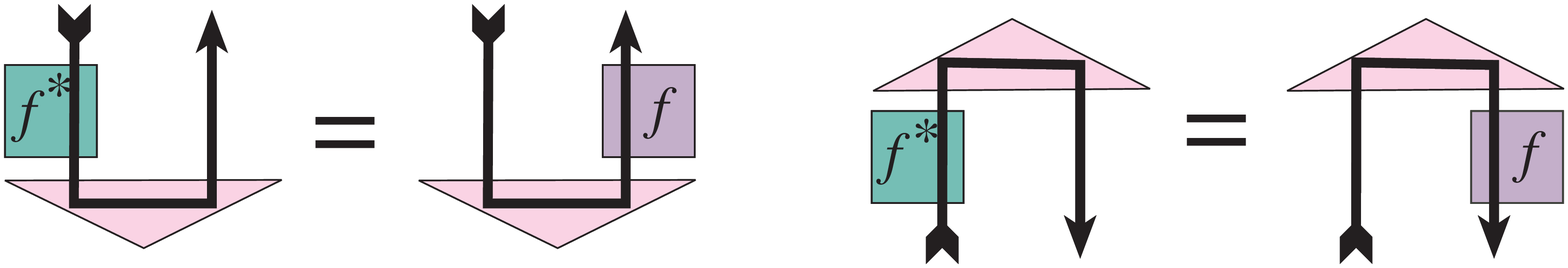,width=360pt}\qquad$\!\!\!$\hfill{\bf (15)}}}  
\end{minipage}
\par\vspace{3mm}\noindent 
and hence we straightforwardly obtain the well-known permutation-law for the trace:
\par\vspace{3mm}\noindent 
\begin{minipage}[b]{1\linewidth}  
\centering{\fbox{\qquad\epsfig{figure=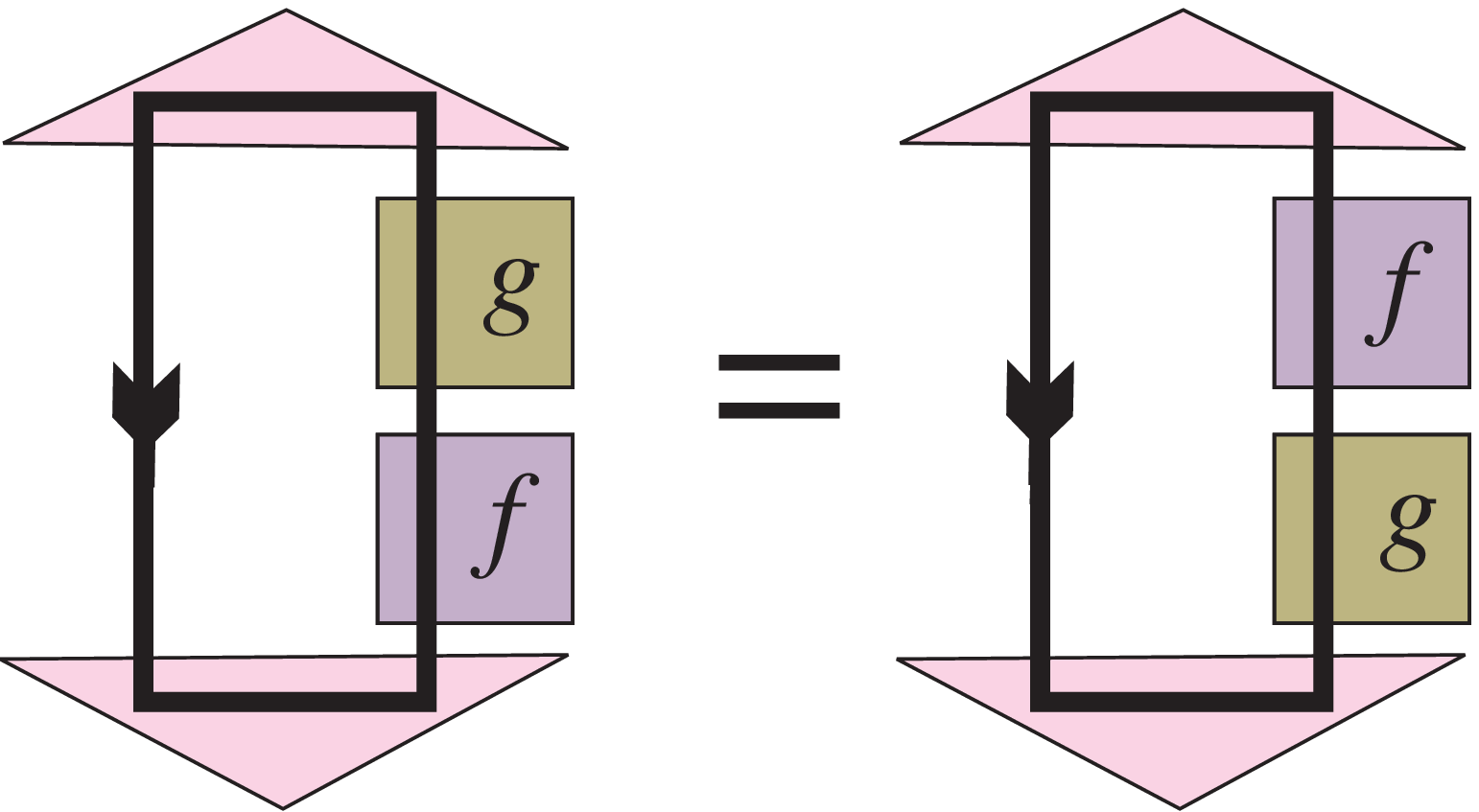,width=160pt}\qquad\hfill{\bf (16)}}}     
\end{minipage}
\par\vspace{3mm}\noindent 
simply by moving either the $f$-box or the $g$-box around the loop while obeying picture {\bf (15)}.
From this follows equality of the two versions of the \em Born-rule \em i.e.~${\sf Tr}(\rho_\phi\circ\PP)=\langle\phi\mid\PP\circ\phi\rangle$ where $\rho_\phi:=|\phi\rangle\langle\phi|$. Indeed,  by picture {\bf (16)} we obtain:
\par\vspace{1mm}\noindent 
\begin{minipage}[b]{1\linewidth}  
\centering{\epsfig{figure=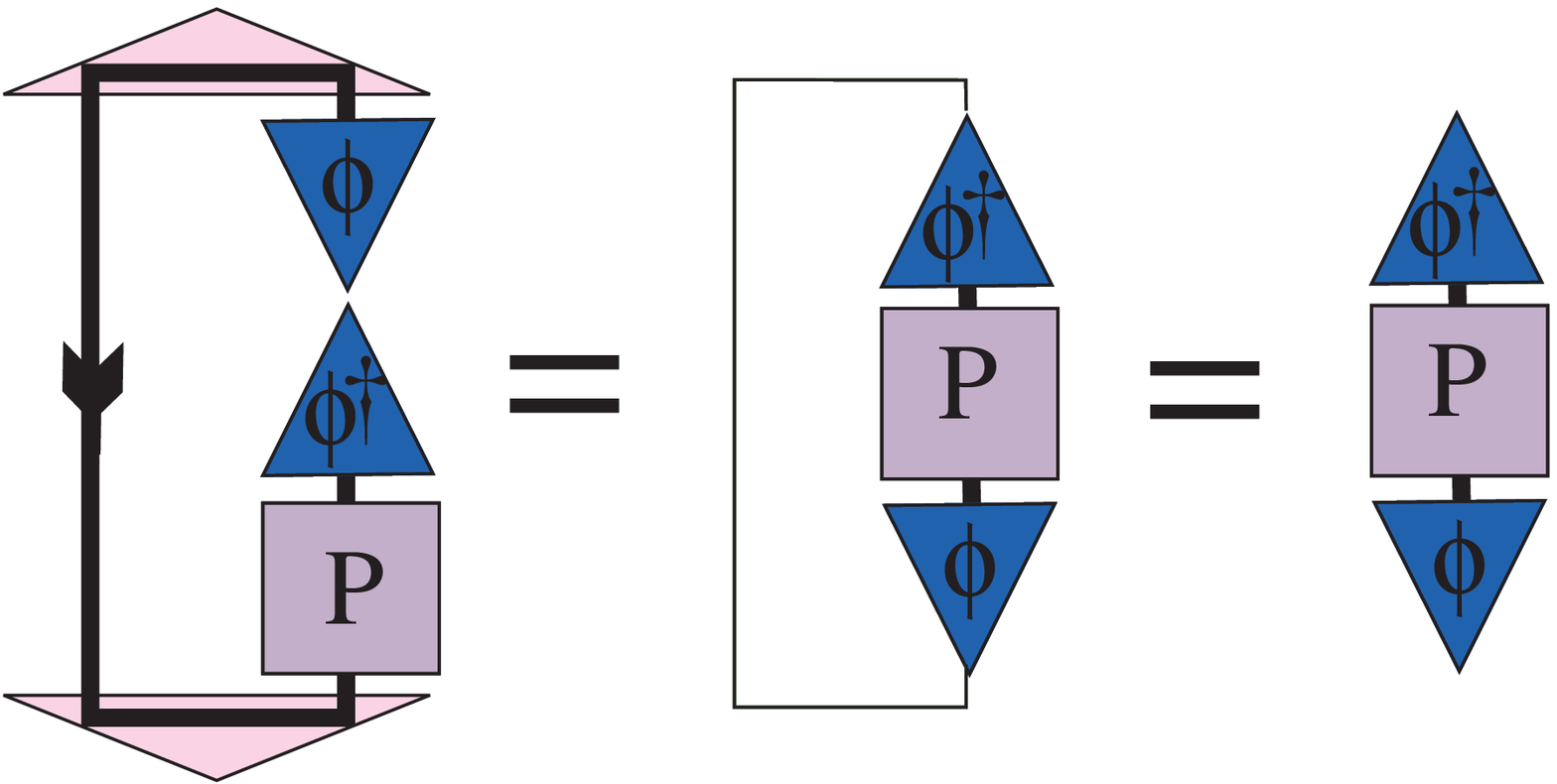,width=220pt}}   
\end{minipage}
\par\vspace{1mm}\noindent 
By the same type of argument we also have a \em Hilbert-Schmidt inner-product \em in our picture calculus \cite{DLL}:
\par\vspace{3mm}\noindent 
\begin{minipage}[b]{1\linewidth}  
\centering{\fbox{\qquad\epsfig{figure=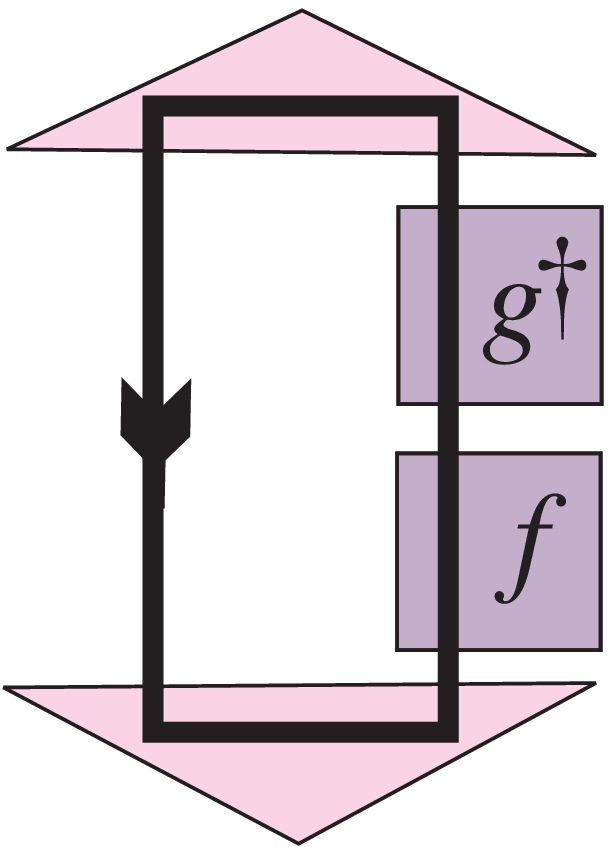,width=70pt}\qquad\hfill{\bf (17)}}}     
\end{minipage}
\par\vspace{3mm}\noindent 
which reduces to the ordinary inner-product for the particular case of states.

\subsection{4.d. Eliminating Global Phases}

Now we will reproduce a quite remarkable result from \cite{DLL}.  It teaches us that if in a picture calculus we pass from an operation to the parallel composition of that operation and its adjoint we \em eliminate redundant global phases\em.  Note that this exactly subsumes the passage from a state vector to the corresponding projector envisioned as a density matrix.  More specifically, what we want to prove is that if:
\par\vspace{1mm}\noindent 
\begin{minipage}[b]{1\linewidth}  
\centering{\epsfig{figure=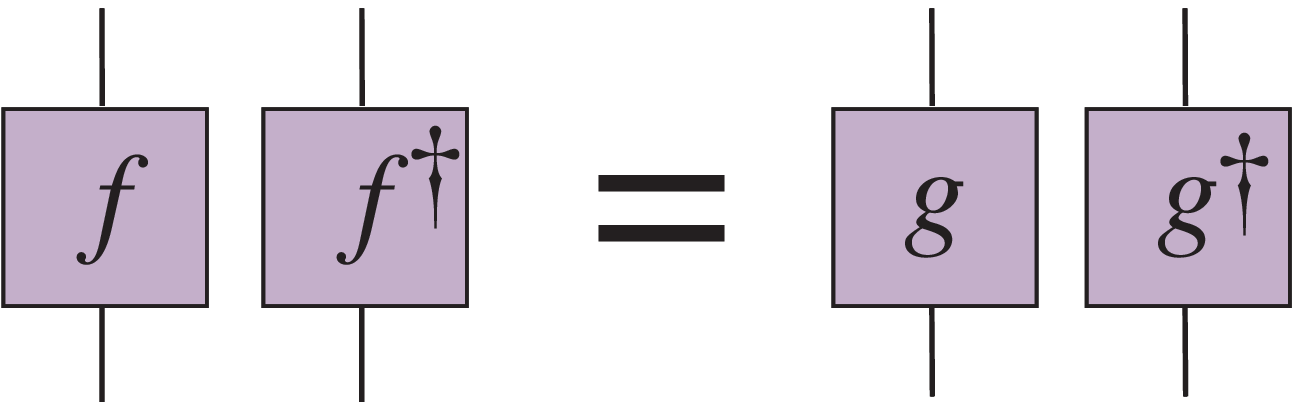,width=140pt}}
\end{minipage}
\par\vspace{-1mm}\noindent 
then there exist diamonds $s,t$ such that:
\par\vspace{1mm}\noindent 
\begin{minipage}[b]{1\linewidth}  
\centering{\epsfig{figure=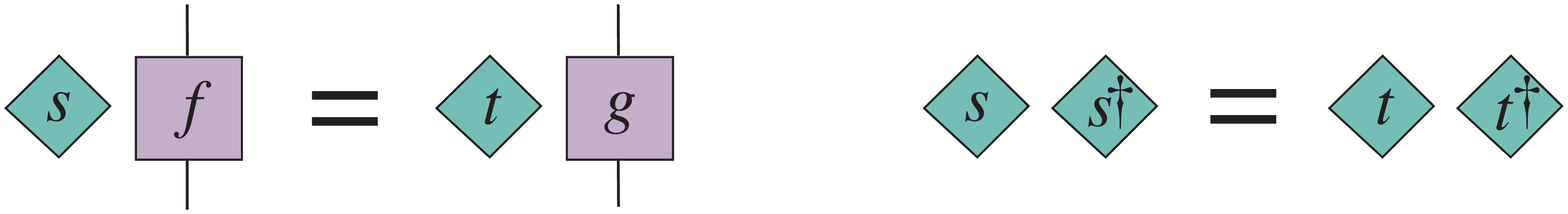,width=320pt}}
\end{minipage}
\par\vspace{1mm}\noindent 
In words, these two equations imply that $f$ and $g$ are equal up to multiplication with numbers of equal lengths and hence  the only difference between $f$ and $g$ is a global phase. Symbolically we can write this statement as:
\[  
f\otimes f^\dagger\!=g\otimes g^\dagger
\ \ \Longrightarrow\ \ 
\exists s,t:\ s\bullet f=t\bullet g\ ,\ s\circ s^\dagger=t\circ t^\dagger
\] 
where the `bullet' denotes scalar multiplication. To prove this we  choose the two diamonds $s,t$ respectively to be two well-chosen particular Hilbert-Schmidt inner-products:
\par\vspace{1mm}\noindent 
\begin{minipage}[b]{1\linewidth}  
\centering{\epsfig{figure=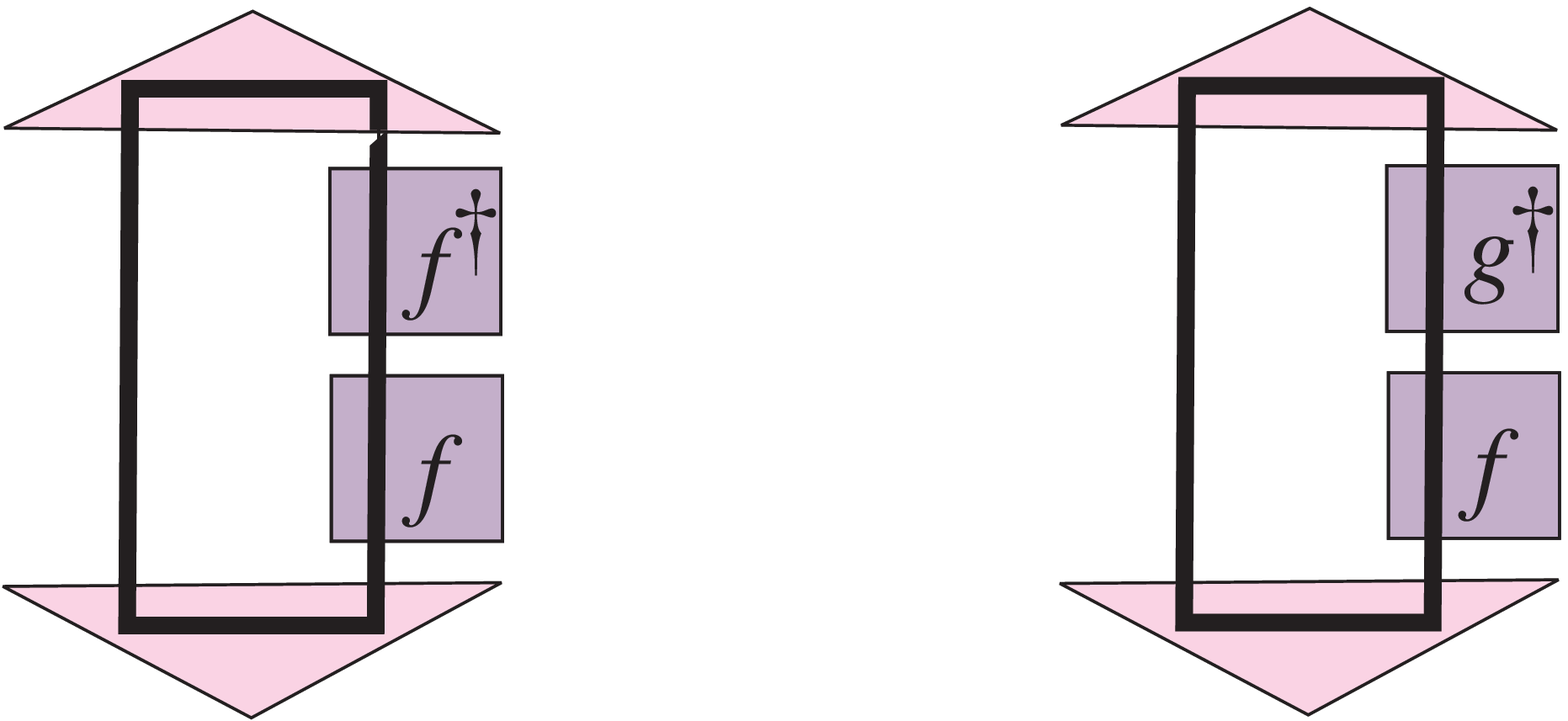,width=210pt}}
\end{minipage}
\par\vspace{1mm}\noindent 
The result now trivially follows by identifying the premiss:
\par\vspace{1mm}\noindent 
\begin{minipage}[b]{1\linewidth}  
\centering{\epsfig{figure=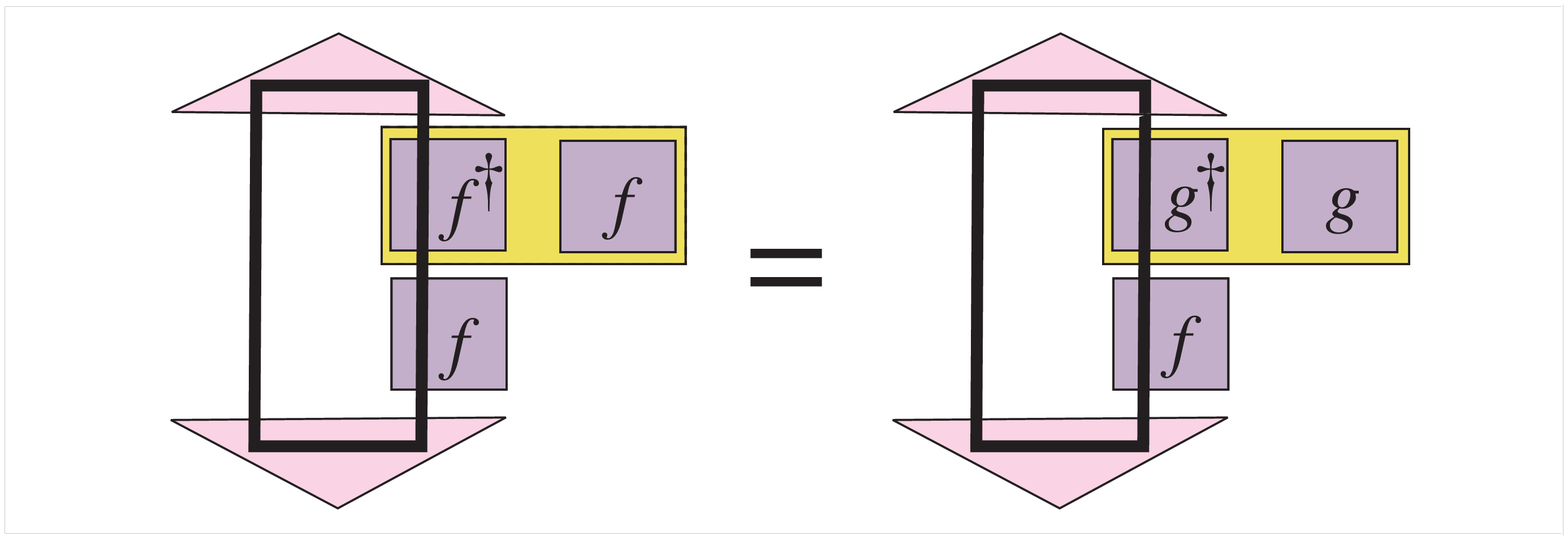,width=304pt}}
\end{minipage}
\par\vspace{1mm}\noindent 
\begin{minipage}[b]{1\linewidth}  
\centering{\epsfig{figure=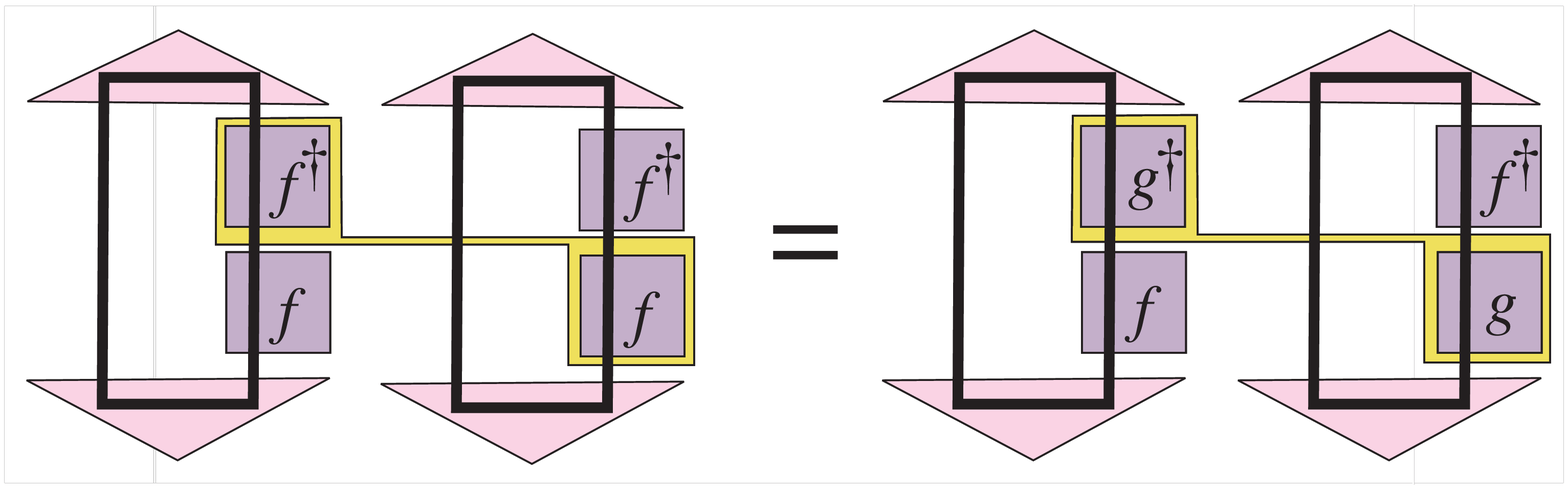,width=342pt}}
\end{minipage}
%\par\vspace{1mm}\noindent 

\subsection{4.e. Completely Positive Maps}

So we now know how to pass to density matrices, and, following Selinger \cite{Selinger}, with this passage to density matrices comes the passage to completely positive maps. Given operations from which we eliminated the global phase:
\par\vspace{-2mm}\noindent 
\begin{minipage}[b]{1\linewidth}  
\centering{\epsfig{figure=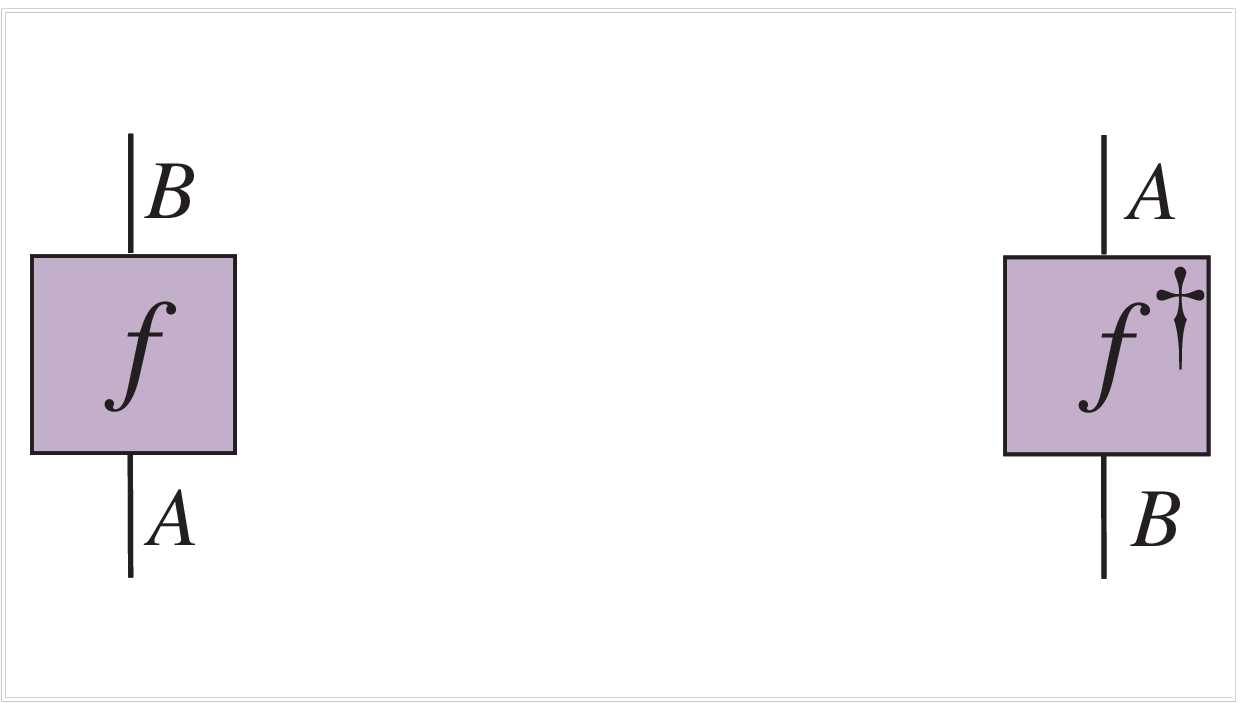,width=140pt}}
\end{minipage}
\par\vspace{-3mm}\noindent 
we allow an ancillary type:
\par\vspace{-3mm}\noindent 
\begin{minipage}[b]{1\linewidth}  
\centering{\epsfig{figure=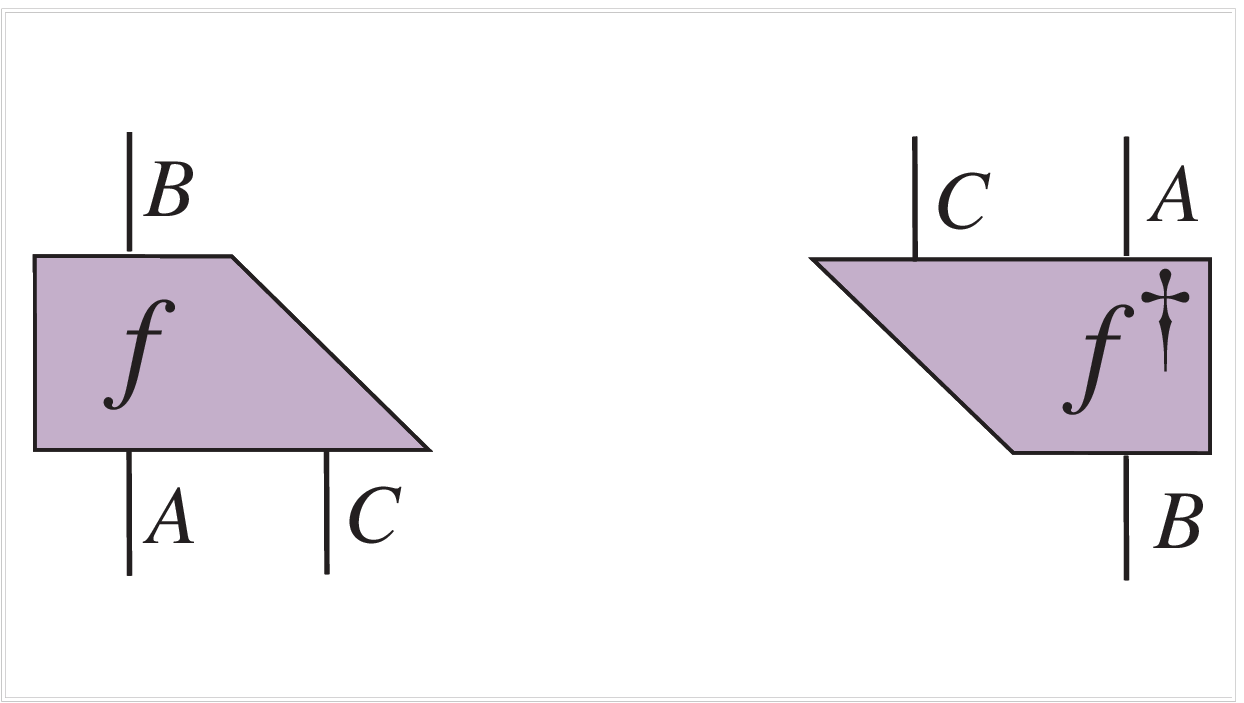,width=140pt}}
\end{minipage}
\par\vspace{-3mm}\noindent 
and when short-circuiting these: 
\par\vspace{3mm}\noindent 
\begin{minipage}[b]{1\linewidth}  
\centering{\fbox{\ \ \ \ \qquad\epsfig{figure=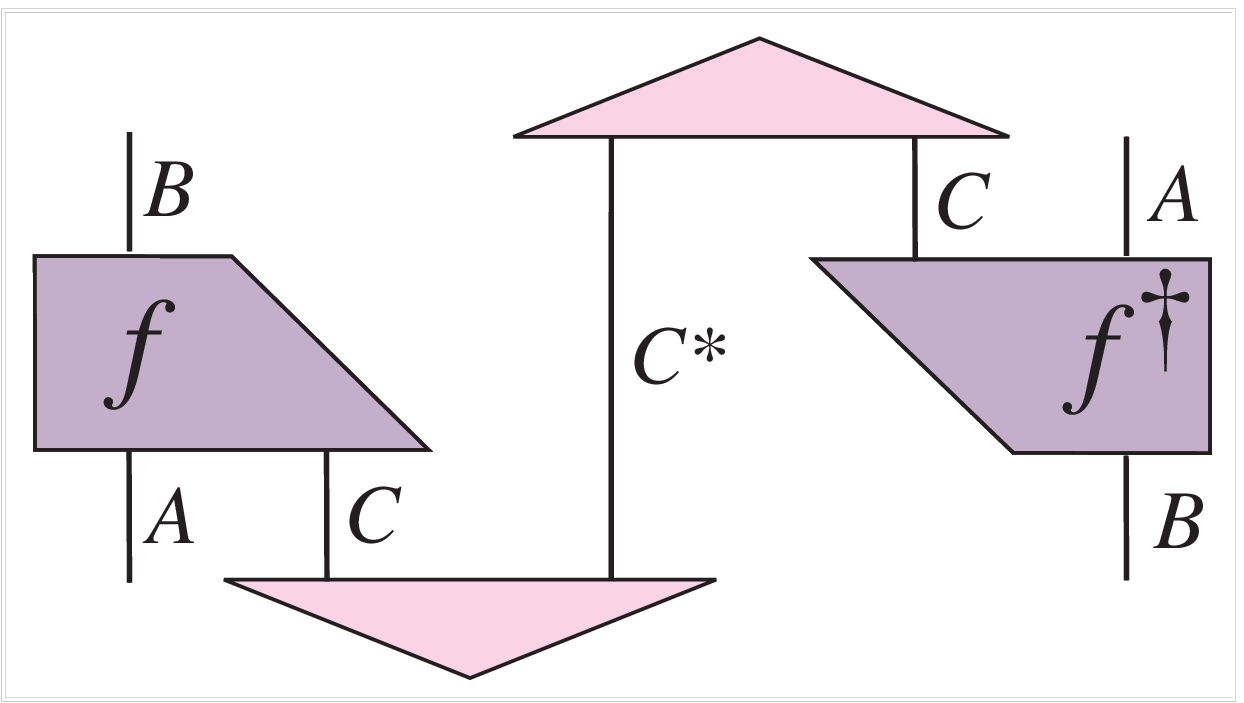,width=140pt}\ \ \quad\hfill{\bf (18)}}}
\end{minipage}
\par\vspace{3mm}\noindent 
it turns out that in the Hilbert space case we exactly obtain all completely positive maps in this manner. Hence the notion of complete positivity is definable for each picture calculus. 
Equivalently (up to some Bell-(co)states), we have:
\par\vspace{3mm}\noindent 
\begin{minipage}[b]{1\linewidth}  
\centering{\fbox{\ \ \ \ \qquad\epsfig{figure=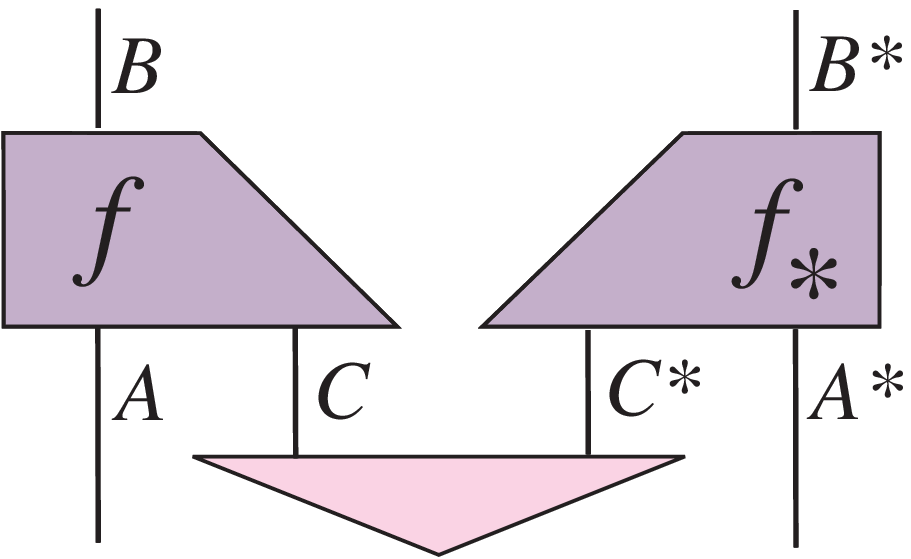,width=109pt}\ \ \quad\hfill{\bf (19)}}}
\end{minipage}
\par\vspace{3mm}\noindent 
providing an alternative representation of completely positive maps which admits covarient composition:
\par\vspace{1mm}\noindent 
\begin{minipage}[b]{1\linewidth}  
\centering{\epsfig{figure=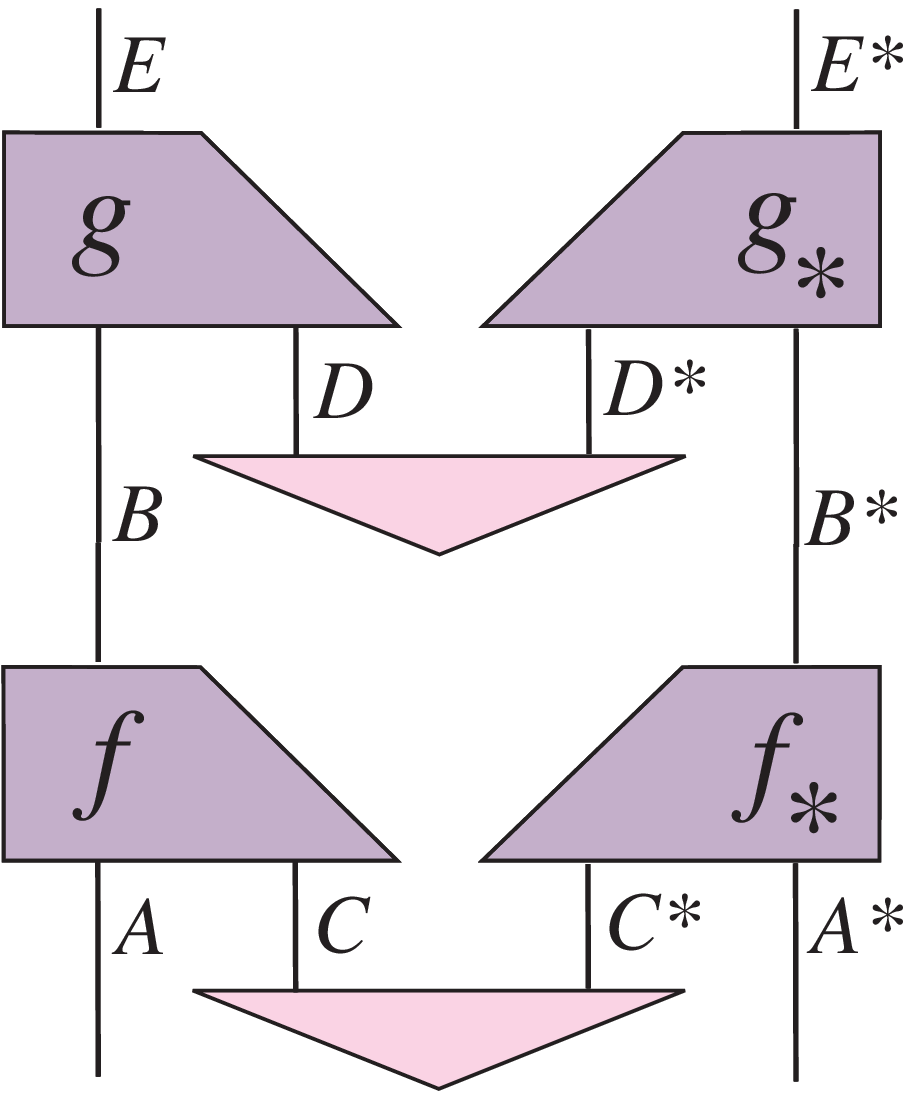,width=109pt}}
\end{minipage}
\par\vspace{1mm}\noindent 
%\section{5. Full categorical quantum mechanics}

\section{5. Literature}

As mentioned in the abstract, the starting point of these developments was the `Logic of Entanglement' \cite{Coe,LN1,LN2}, which emerged from an investigation on the connection between Quantum Entanglement and Geometry of Interaction \cite{PT}, and which provided a scheme to derive protocols such as Quantum Teleportation \cite{BBC}, Logic-gate Teleportation \cite{Gottesman}, Entanglement Swapping \cite{Swap} and various related ones through the notion of `Quantum Information-flow' (see also \cite{LN1,LN2}).  In \cite{AC1,AC1.5} Abramsky and I axiomatized this 
quantum information-flow in category theoretic terms which via the work of Kelly \& Laplaza \cite{KellyLaplaza} and Joyal \& Street \cite{JoyalStreet} formally justifies the grapical calculus informally initiated in \cite{Coe}, subsequently refined in \cite{AC1,LN1,AC1.5}, and
connected up by Abramsky \& Duncan to the so-called proof nets of Linear Logic \cite{AbramskyDuncan}.
The most precise account on this issue can currently be found in Selinger's paper \cite{Selinger} --- while at first sight his graphical calculus looks quite different from ours, they are in fact equivalent, being both transcriptions of strong compact closure (or as Selinger calls it, dagger compact closure).  Actually, the use of graphical calculi for tensor calculus goes back to Penrose \cite{Penrose}, initiating as applications the theories of Braids and Knots in Mathematical Physics.  We mention independent work by Louis Kauffman \cite{Kauffman} which provides a topological interpretation of quantum teleportation and hence relates to the Logic of Entanglement in \cite{Coe}, and we mention independent work by John Baez \cite{Baez} which relates to the developments in \cite{AC1,AC1.5} in the sense that he exposes similarities of the category of relations, the category of Hilbert spaces, and the compact closed category of cobordisms which plays an important role in Topological Quantum Field Theory. There also seem to be promising connections with Basil Hiley's recent work \cite{Hiley} on Dirac's `standard ket' \cite{Dirac} in the context of quantum evolution. We also have 
\begin{center}
Strong Compact Closure \cite{AC1,AC1.5} $\Rightarrow$ Kelly's Compact Closure \cite{Kelly} $\Rightarrow$ Barr's $*$-autonomy \cite{Barr} 
\end{center}
where the latter is the semantics for the multiplicative fragment of Linear Logic \cite{Seely}.  Linear Logic itself is a logic in which one is not allowed to copy nor delete premisses, hence enabling one to take computational resources into account --- in view of the No-Cloning and No-Deleting theorems it is not a surprise that the axiomatization of quantum information-flow comprises this resource-sensitive logicality.  A very substantial contribution to our program was made by Peter Selinger who discovered the construction which turns any strongly compact closed category of pure states and pure operations into one of mixed states and completely positive maps \cite{Selinger}. At the same time, I discovered the preparation-state agreement axiom \cite{DLL}, and in currently ongoing work I identified another axiom which combines preparation-state agreement and the structural content of Selinger's construction. It has the potential 
to  provide a categorical foundation for quantum information theory \cite{QI} --- many quantum information theoretic fidelities and capacities (see \cite{Werner} for a structured survey on some of these) can indeed be unified in our graphical calculus once this axiom is added. At the more abstract side of the spectrum we refer for an abstract theory of partial trace to  \cite{JSV} and for free constructions of traced and strongly compact closed categories to \cite{Abr}.

\section{Acknowledgments}
 
Feed-back on my talks by Samson Abramsky, Dan Browne, Vincent Danos, Basil Hiley, Louis Kauffman, Prakash Panangaden, Mehrnoosh Sadrzadeh and Peter Selinger has been very useful when preparing these notes. The research which led to these results was
supported by the EPSRC grant EP/C500032/1 on High-Level Methods in
Quantum Computation and Quantum Information. We are greatfull to Andrei Khennikov at Vaxjo University, Dusko Pavlovic at Kestrel Institute, Google, Perimeter Institute and Texas A\&M University for the invitations to present this work.

\end{document}